\documentclass[twocol]{ametsoc}

\usepackage{text_shortcuts}
\usepackage{math_shortcuts}
\usepackage{phys_shortcuts}

\usepackage[varg]{newtxmath}  % makes \nu and v look less similar

\journal{jas}
\bibpunct{(}{)}{;}{a}{}{,}

\title{Axisymmetric constraints on cross-equatorial Hadley cell extent}

\authors{Spencer A. Hill\correspondingauthor{Spencer Hill, UCLA Atmospheric and Oceanic Sciences, Box 951565, Los Angeles, CA 90095-1565}}
\affiliation{Department of Atmospheric and Oceanic Sciences, University of California, Los Angeles and Division of Geological and Planetary Sciences, California Institute of Technology, Pasadena, California}
\email{shill@atmos.ucla.edu}

\extraauthor{Simona Bordoni}
\extraaffil{Division of Geological and Planetary Sciences, California Institute of Technology, Pasadena, California}
\extraauthor{Jonathan L. Mitchell}
\extraaffil{Department of Earth, Planetary, and Space Sciences, and Department of Atmospheric and Oceanic Sciences, University of California, Los Angeles}

\abstract{We consider the relevance of known constraints from each of Hide's theorem, the angular momentum conserving (AMC) model, and the equal-area model on the extent of cross-equatorial Hadley cells.  These theories respectively posit that a Hadley circulation must span: all latitudes where the radiative convective equilibrium (RCE) absolute angular momentum (\(M\rce\)) satisfies \(M\rce>\Omega a^2\) or \(M\rce<0\) or where the RCE absolute vorticity (\(\etarce\)) satisfies \(f\etarce<0\); all latitudes where the RCE zonal wind exceeds the AMC zonal wind; and over a range such that depth-averaged potential temperature is continuous and that energy is conserved.  The AMC model requires knowledge of the ascent latitude \(\ascentlat\), which need not equal the RCE forcing maximum latitude \(\maxlat\).  Whatever the value of \(\ascentlat\), we demonstrate that an AMC cell must extend at least as far into the winter hemisphere as the summer hemisphere.  The equal-area model predicts \(\ascentlat\), always placing it poleward of \(\maxlat\).  As \(\maxlat\) is moved poleward (at a given thermal Rossby number), the equal-area predicted Hadley circulation becomes implausibly large, while both \(\maxlat\) and \(\ascentlat\) become increasingly displaced poleward of the minimal cell extent based on Hide's theorem (\ie/ of supercritical forcing).  In an idealized dry general circulation model, cross-equatorial Hadley cells are generated, some spanning nearly pole-to-pole.  All homogenize angular momentum imperfectly, are roughly symmetric in extent about the equator, and appear in extent controlled by the span of supercritical forcing.
}

\begin{document}
\maketitle

\section{Introduction}
Except during equinox, insolation always has a nonzero meridional derivative spanning the equator, precluding a state of latitude-by-latitude radiative convective equilibrium (RCE).  The resulting thermal- or gradient-balanced wind would asymptote to \(+\infty\) on the winter side of the equator and \(-\infty\) on the summer side.  An overturning circulation must emerge that removes this physically impossible feature by redistributing heat and angular momentum.  The resulting solsticial cross-equatorial Hadley cells in our solar system, however, differ dramatically in scale --- ascending at relatively low latitudes in the summer hemisphere on Earth \vs/ nearly at the summer pole on Venus \citep{gierasch_meridional_1975} and Titan \citep{mitchell_climate_2016}.  Nevertheless, each spans at least as far into the winter hemisphere as the summer hemisphere.  This paper seeks minimal, prognostic, qualitatively accurate theoretical arguments that account for these features.

The inviability of gradient balance at the equator is one manifestation of the well-known Hide's theorem --- a set of conditions determining if the distributions of absolute angular momentum (\(\Mrce\)) and absolute vorticity (\(\etarce\)) in the hypothetical RCE state are physically realizable \citep{hide_dynamics_1969,schneider_axially_1977}.  Any latitude where the RCE state violates the conditions of Hide's theorem is said to be supercritically forced, and an overturning circulation must span at minimum all supercritical latitudes.  Of particular note is the condition (expanded upon below) that the RCE absolute vorticity cannot take the opposite sign of the local Coriolis parameter: \(f\etarce<0\) is forbidden \citep{plumb_response_1992,emanuel_thermally_1995}, making the poleward extent of supercritical forcing in the summer hemisphere bounded by the latitude where \(\etarce=0\) in most cases.

In axisymmetric atmospheres in which viscosity is weak above the boundary layer, the cross-equatorial Hadley cell is usefully described by the angular momentum conserving (AMC) model.  In this framework, ascent concentrated at a single latitude (\(\ascentlat\)) imparts that latitude's planetary angular momentum to the free troposphere, which the Hadley circulation then homogenizes throughout its confines \citep{held_nonlinear_1980,lindzen_hadley_1988}; \(\ascentlat\) is also the boundary separating the cross-equatorial cell and the smaller, summer cell.  To satisfy Hide's theorem, such an AMC circulation must span all latitudes wherein the angular momentum value of the RCE state exceeds the AMC value.

A well-known theory for \(\ascentlat\) in an AMC Hadley circulation comes from the equal-area model, which assumes Newtonian relaxation of temperatures toward a specified RCE distribution, continuity of column-integrated temperature at each cell edge, and conservation of energy integrated over each cell.  Given these, it predicts \(\ascentlat\) as well as the poleward edges of both the cross-equatorial winter cell and the summer cell and the column-integrated temperature at \(\ascentlat\) \citep{held_nonlinear_1980,lindzen_hadley_1988}.  The equal-area solution for the cross-equatorial Hadley cell grows rapidly as the latitude of the RCE thermal maximum, \(\maxlat\), is moved poleward.  For the largest value shown by \citet{lindzen_hadley_1988}, \(\maxlat\approx8^\circ\), the cross-equatorial cell spans \({\sim}45^\circ\)S-28\degr{}N or \({\sim}41^\circ\)S-23\degr{}N depending on the value of the imposed fractional meridional equilibrium temperature drop factor (\(\Delta_\mr{h}=1/3\) or 1/6, respectively, c.f. their Fig.~4).  In addition, it always predicts \(\ascentlat\geq\maxlat\) and is agnostic to the extent of supercritical forcing.

Yet, under solsticial forcing, insolation maximizes at the summer pole, and thus the effective RCE \(\maxlat\) should also.  \citet{faulk_effects_2017} and \citet{singh_limits_2019} perform perpetual solsticial forcing simulations in an idealized aquaplanet GCM, finding that the resulting cross-equatorial cell is confined to the Tropics, as on Earth, unless the rotation rate is decreased.  These studies also demonstrate that the latitude at which \(\etarce=0\) predicts \(\ascentlat\) with qualitative accuracy both at Earth's rotation rate \citep{faulk_effects_2017} and as the rotation rate is varied over a wide range \citep{singh_limits_2019}.  Moreover, while the traditional AMC model dictates that the circulation and column-integrated equivalent potential temperature (\(\hatthetae\)) fields mutually evolve such that \(\ascentlat\) coincides with a local maximum in \(\hatthetae\) \citep{lindzen_hadley_1988,emanuel_thermally_1995,prive_monsoon_2007}, in the Faulk et al.\@ and Singh simulations the near-surface moist static energy (a good indicator of \(\hatthetae\)) always maximizes at the summer pole --- nearly a hemisphere away from \(\ascentlat\) in the highest rotation rate cases.

In what follows, we demonstrate using a dry, axisymmetric GCM that the \(\etarce=0\) latitude predicts \(\ascentlat\) with qualitative accuracy --- and the equal-area model does not --- both under conventional forcings and in exotically forced cases that generate planetary scale Hadley circulations while remaining at Earth's rotation rate.  We also show without appeal to the equal-area model that a cross-equatorial AMC cell must extend as far or farther into the winter hemisphere as into the summer hemisphere.  Hide's theorem is reviewed and its various previous manifestations synthesized in Section~\ref{sec:hides-theorem}.   Section~\ref{sec:extent} reviews AMC theory and the equal-area model and compares their predictions for the cross-equatorial cell's edges with those stemming directly from Hide's theorem.   Section~\ref{sec:crit-sims} presents the results of the numerical simulations, after which we conclude with a summary and discussion in Sections~\ref{sec:summ} and \ref{sec:disc}, respectively.  We heavily utilize the results of \citet{held_nonlinear_1980}, \citet{lindzen_hadley_1988}, \citet{plumb_response_1992}, and \citet{emanuel_thermally_1995} and refer to them henceforth as HH80, LH88, PH92, and E95, respectively.

Before proceeding, we note that, even in axisymmetric cases where the complicating factors of eddy stresses can be neglected, simulated Hadley cells never truly approach the AMC limit \citep{held_nonlinear_1980,lindzen_hadley_1988,adam_global_2009}.  In eddying atmospheres, the AMC assumption becomes even more problematic, although in observations and simulations the solsticial, cross-equatorial cell is more nearly AMC than are the summer, equinoctial, or annual-mean cells \citep[\eg/][]{schneider_general_2006}.   We expand upon these and other caveats, including the complicating effects of moisture, in concluding subsections within Sections~\ref{sec:hides-theorem} and \ref{sec:extent}.  Additional subtleties discussed in footnotes and the appendices may be skipped by casual readers.  We otherwise proceed using the original dry, axisymmetric framework.

\section{Hide's theorem}
\label{sec:hides-theorem}
After presenting the governing equations, this section reviews fundamental properties of the gradient-balanced RCE state, synthesizes the various forms Hide's theorem has taken in past literature, and notes some important caveats.

We consider dry, axisymmetric, Boussinesq atmospheres under time invariant radiative forcing, with radiative transfer represented as Newtonian cooling of potential temperature \(\theta\) toward an RCE potential temperature field \(\thetarce\) that is known analytically.  The corresponding governing equations are
\begin{subequations}
\begin{align}
  \label{eq:bouss-eqs}
  0&=-\uv\cn u+fv+\dfrac{uv\tan\lat}{a}+\pd{}{z}\left(\nu\pd{u}{z}\right)\\
  0&=-\uv\cn v -fu+\dfrac{u^2\tan\lat}{a}-\dfrac{1}{a}\pdlat{\Phi}+\pd{}{z}\left(\nu\pd{v}{z}\right)\\
  0&=-\uv\cn\theta-\dfrac{\theta-\thetarce}{\tau}+\pd{}{z}\left(\nu\pd{\theta}{z}\right)\\
  0&=-\nc\mb{v}\\
  0&=\pd{\Phi}{z}-g\dfrac{\theta}{\theta_0}.
\end{align}
\end{subequations}
Here, \(\uv=(v,w)\) with \(v\) the meridional velocity and \(w\) the vertical velocity, \(u\) is zonal velocity, \(f=2\Omega\sinlat\) is the Coriolis parameter with \(\Omega\) the planetary rotation rate and \(\lat\) latitude, \(a\) is planetary radius, \(\nu\) is the kinematic viscosity, \(\Phi=gz\) is geopotential height, \(\tau\) is the Newtonian cooling timescale, \(\theta_0\) is the Boussinesq reference potential temperature, and other terms have their standard meaning.  Note that the Boussinesq equations are isomorphic to the fully compressible equations in pressure coordinates \citep[\eg/][]{vallis_atmospheric_2017}.

\subsection{Gradient wind balance in radiative convective equilibrium}
By definition, latitude-by-latitude RCE requires \(v=w=0\), in which case combining the meridional momentum and hydrostatic balance equations (1b) and (1e) leads to RCE zonal wind and potential temperatures (\(\urce\) and \(\theta\rce\), respectively) in gradient wind balance:
\begin{subequations}
\begin{equation}
  \label{eq:grad-wind}
  \pd{}{z}\left(\dfrac{\tan\lat}{a}\urce^2+f\urce\right)=-\dfrac{g}{a\theta_0}\pdlat{\theta\rce}.
\end{equation}
Assuming drag in the boundary layer is large enough that the surface zonal wind is negligible\footnote{More formally, assuming lower boundary conditions of \(\nu\pdsl{u}{z}=Cu\), \(\nu\pdsl{v}{z}=Cv\), where \(C\) is a constant drag coefficient, \cf/ Eq.~3 of HH80.}, the integral of \eqref{eq:grad-wind} from the surface to some height \(z\) yields
\begin{equation}
  \label{eq:grad-wind-int}
  \dfrac{\tan\lat}{a}\urce^2(z)+f\urce(z)+\dfrac{gz}{a\theta_0}\pdlat{\hat{\theta}\rce}=0,
\end{equation}
\end{subequations}
where \(\hat{\theta}\) is the average of \(\theta\) between the surface and \(z\).  \eqref{eq:grad-wind-int} is a quadratic equation for \(\urce\) that can be solved directly.  Choosing the positive root that corresponds to \(\urce=0\) at the surface as required, this is
\begin{subequations}
\begin{equation}
  \label{eq:u-rce}
  \urce=\Omega a\coslat\left[\sqrt{1-\frac{1}{\coslat\sinlat}\frac{gz}{\Omega^2a^2\theta_0}\pdlat{\hat{\theta}\rce}}-1\right].
\end{equation}
Away from the equator, if \(\pdsl{\hat{\theta}\rce}{\lat}=0\) then \(\urce=0\).  At the equator, if \(\pdsl{\hat{\theta}\rce}{\lat}=0\), then \(\lim_{\lat\rightarrow0}\urce\) is indeterminate.  L'H\^opital's rule then gives
\begin{equation}
  \label{eq:u-rce-lhopital}
  \urce(\phi{=}0, \pdsl{\hat{\theta}\rce}{\lat}{=}0) =\Omega a\left[\sqrt{1-\dfrac{gz}{\Omega^2a^2\theta_0}\dfrac{\partial^2\hat{\theta}\rce}{\partial\lat^2}}-1\right].
\end{equation}
\end{subequations}
Accordingly, any equatorial maximum in potential temperature with nonzero second derivative generates equatorial westerlies.\footnote{If the second derivative is also zero, zonal flow at the equator will remain zero (as is the case for the AMC solutions described below), but this is unlikely for the RCE state in the annual mean for Earth-like orbits \citep[as argued more formally by][]{schneider_general_2006}.  Flatter annual-mean profiles become more relevant for planets with larger orbital obliquities \citep[with the poles receiving more annual mean insolation than the equator for orbits with obliquities \({\gtrsim}55^\circ\); \eg/][]{linsenmeier_climate_2015}, although in those cases the annual mean Hadley cells are likely the small residual of very strong, seasonally reversing cells that rarely approach cross-equatorial symmetry.}

\subsection{Hide's theorem: existing forms and synthesis}
Provided \(\nu\neq0\) and that the effect of transients is negligible, in steady state the zonal momentum equation (1a) may be written
\begin{equation}
  \label{eq:abs_ang_mom_budg}
  \uv\cn M=\pd{}{z}\left(\nu\pd{M}{z}\right),
\end{equation}
where \(M=a\coslat(\Omega a\coslat+u)\) is absolute angular momentum per unit mass, and recall that all values are steady-state averages.  \eqref{eq:abs_ang_mom_budg} precludes isolated extrema in \(\ol{M}\).  Mathematically, at any such an extremum, \(\nabla\ol{M}=\mathbf{0}\), so that the left-hand side of \eqref{eq:abs_ang_mom_budg} must vanish but not the right-hand side.  Physically, viscous diffusion that acts to flatten out the extremum would have to be balanced by momentum flux convergence for a maximum (divergence for a minimum), which would require time-mean mass convergence (divergence for a minimum), violating conservation of mass (see Appendix A of PH92 for a formal proof).  Only at the surface, where frictional stress can balance the diffusive term, can an extremum occur.  In particular, the extremal values of planetary angular momentum (\(M=\Omega a^2\) at the equator and \(M=0\) at either pole) must bound \(M\) at all latitudes.

Early expressions of Hide's theorem amount to the application of this result to maxima in the angular momentum field of the hypothetical RCE state, \(\Mrce\); namely, if \(\Mrce>\Omega a^2\) away from the surface at any latitude, a Hadley circulation must emerge \citep{hide_dynamics_1969,schneider_axially_1977}.  It follows from \eqref{eq:u-rce} and \eqref{eq:u-rce-lhopital} that any nonzero first or second meridional derivative in \(\hatthetarce\) at the equator is supercritical.  By the same arguments applied to a angular momentum minima, \(\Mrce<0\) is also forbidden, but interestingly negative angular momentum values are also forbidden for another reason: a real-valued solution to \eqref{eq:u-rce} does not exist if the quantity within the square root operator is negative, and the minimum real solution is \(\urce=-\Omega a\coslat\), which yields \(M\rce=0\) at all latitudes \citep{fang_simple_1996,adam_global_2009}.

Away from the equator, local absolute angular momentum extrema are readily identified from the absolute vorticity distribution.  The meridional derivative of absolute angular momentum is proportional to the vertical component of absolute vorticity \(\eta\).  Specifically, \(\pdsl{M}{\lat}=-(a^2\coslat)\eta\), where \(\eta=f+\zeta\), with \(\zeta=-(a\coslat)^{-1}\pdsl{(u\coslat)}{\lat}\) the relative vorticity (hereafter, \(\eta\) is referred to without confusion as absolute vorticity).  Accordingly, \(\etarce=0\) at any local extremum in \(\Mrce\).  Because \(\eta=f\) in the absence of flow, this vorticity-based sufficient condition for supercriticality may be compactly expressed as \(f\etarce<0\) (PH92, E95).

\begin{figure}[t]
  \centering
  \noindent\includegraphics[width=19pc,angle=0, trim={0 0cm 0 0.5cm}, clip]{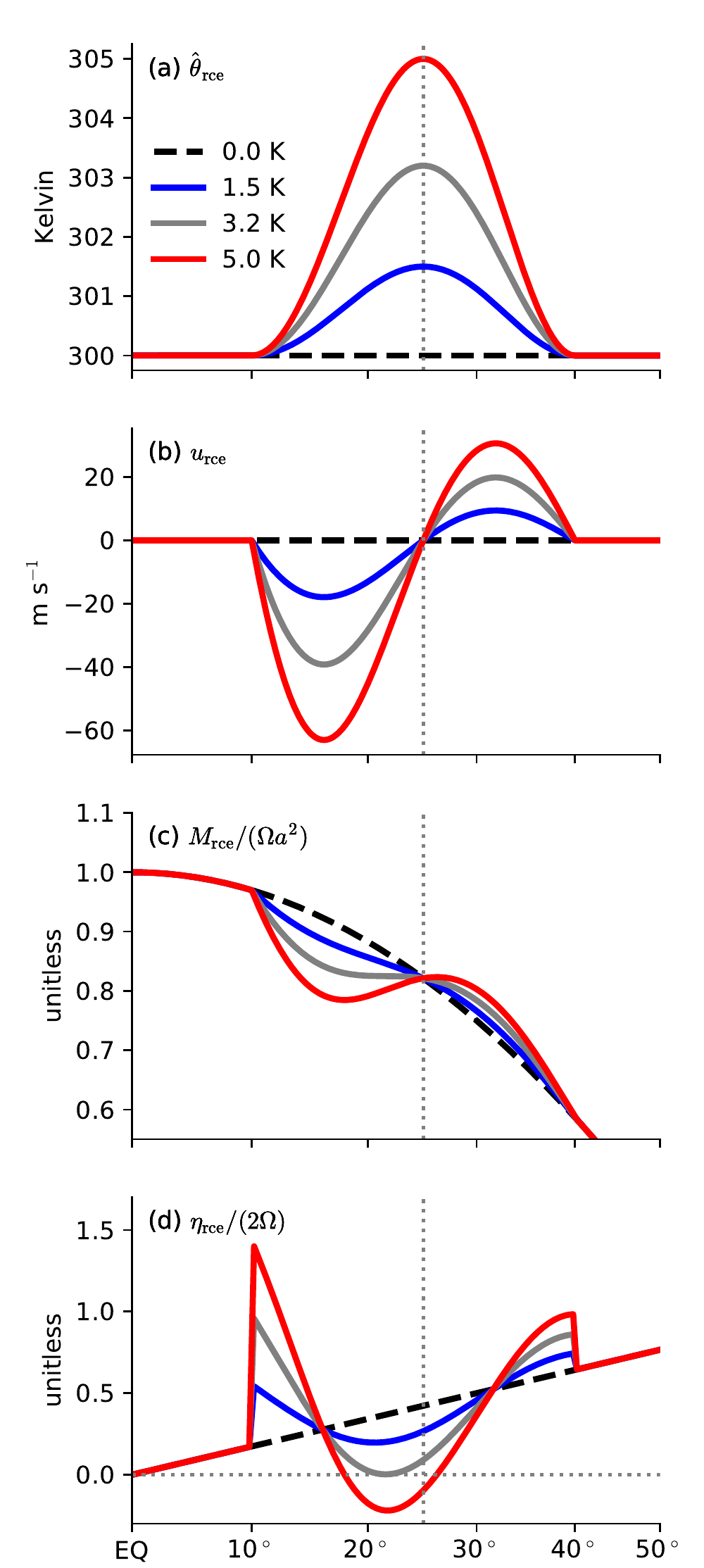}\\
  \caption{RCE profiles corresponding to the forcing used by \citet[][their Eq.~9]{plumb_response_1992}, which comprises uniform \(\hat\theta\rce\) at all latitudes other than a ``bump'' centered on 25\degr{}N (indicated by the vertical gray dotted line) and 30\degr{} wide, of (a) column-averaged potential temperature (in~Kelvin), (b) zonal wind (in \ms/), (c) absolute angular momentum normalized by the planetary angular momentum at the equator, and (d) absolute vorticity normalized by twice the planetary rotation rate, for different magnitudes of the forcing maximum as indicated by the legend in (a).  The horizontal dotted gray line in (d) marks the zero line.}
  \label{fig:ph92-schematic}
\end{figure}

Figure~\ref{fig:ph92-schematic} visualizes this \(f\etarce<0\) condition by showing the potential temperature, zonal wind, angular momentum, and absolute vorticity fields corresponding to the depth-averaged forcing of PH92 (their Eq.~9), which comprises uniform \(\hatthetarce\) everywhere except for a local maximum centered at 25\degr{}N dropping off as \(\cos^2\lat\) in a 30\degr{} wide region, for successively larger values of the forcing maximum.  With no maximum present (dashed black curves), \(\urce=0\) everywhere, and \(M\rce\) and \(\etarce\) take their planetary values.  Introducing a weak forcing maximum (blue curves) generates easterlies on the maximum's equatorward side and westerlies on the poleward side.  The easterlies bend down, and the westerlies bend up, the \(M\rce\) curve, but not enough to generate any extrema in \(M\rce\): \(\etarce\) retains its original sign everywhere, and the forcing is subcritical.  Increasing the magnitude of the forcing maximum causes the easterlies and westerlies to intensify, eventually enough to generate \(\etarce=0\) at a point slightly equatorward of \(\maxlat\) (gray curves).  A forcing maximum that is any stronger is supercritical (red curves).  \(M\rce\) develops a minimum equatorward and a maximum poleward of \(\maxlat\), between which \(\etarce\) has changed sign.

E95 shows that the \(f\etarce<0\) condition also applies at the tropopause in non-axisymmetric and/or purely inviscid atmospheres, the derivation of which is provided in Appendix A along with two additional physical interpretations.

The \(f\etarce<0\) condition may be expressed in terms of \(\hatthetarce\) as, \cf/ PH92,
\begin{equation}
  \label{eq:ph92-crit}
  C_{\mr{ph92}}=4\Omega^2a^2\cos^3\lat\sinlat + \frac{gH}{\theta_0}\pdlat{}\left(\frac{\cos^3\lat}{\sinlat}\pdlat{\hatthetarce}\right).
\end{equation}
\(C_{\mr{ph92}}>0\) corresponds to supercritical forcing in the Northern Hemisphere and subcritical forcing in the Southern Hemisphere (signs that are the same as for \(\etarce\) itself).  Note that this can be re-expressed in terms of the planetary Burger number \(gH/\Omega^2a^2\) by dividing both sides by \((\Omega a)^2\).

In summary, any latitude in an axisymmetric atmosphere with nonzero free tropospheric viscosity that satisfies any of the following three conditions is supercritical:
\begin{enumerate}
  \item \(M\rce>\Omega a^2\) (global maximum in \(M\rce\))
  \item \(M\rce<0\) (global minimum in \(M\rce\) and complex-valued \(\urce\))
  \item \(f\etarce<0\) (local extrema in \(M\rce\) and unrealizable sign change in \(\etarce\)).
\end{enumerate}
And Hide's theorem states that the existence of any supercritical latitude makes the RCE state physically impossible, meaning that a Hadley circulation must emerge.

\subsection{Caveats}
We are utilizing the thin-shell limit, wherein vertical variations in the moment arm are taken as negligible compared to meridional variations.  This is appropriate for terrestrial bodies but not the gas giants.  See \eg/ \citet{oneill_slantwise_2016} for consideration of angular momentum dynamics in deep atmospheres.

Neglected in \eqref{eq:abs_ang_mom_budg} is the divergence of eddy momentum fluxes, \(\nc\ol{M'\uv'}\), where primes denote deviations from the time-mean and the overbar a temporal average.  In non-axisymmetric atmospheres, zonally asymmetric eddies can generate interior extrema in the angular momentum field, the most notorious example being a westerly, ``superrotating'' jet in the equatorial troposphere, through a variety of mechanisms \citep[\eg/][]{schneider_formation_2009,caballero_spontaneous_2010,mitchell_transition_2010,wang_planetary_2014}.  In simulations of axisymmetric atmospheres (including those we present in Section~\ref{sec:crit-sims}), propagating symmetric instabilities are ubiquitous \citep[\eg/][]{satoh_hadley_1994} and can, in principle, effect nontrivial momentum flux divergences.

Another striking example of the influence of eddy momentum flux divergences is the emergence of Hadley cells in simulations with uniform insolation and other boundary conditions, which by the axisymmetric arguments we presented should have \(\urce=0\) everywhere and thus no Hadley circulation; instead, transient eddies (both in axisymmetric and zonally varying simulations) transport momentum meridionally, necessitating compensating Hadley cell angular momentum transports to attain a balanced budget \citep[\eg/][]{kirtman_spontaneously_2000,shi_large-scale_2014,merlis_surface_2016}.

We lack at present a satisfying explanation for the coincidence of the \(M\rce<0\) manifestation of Hide's theorem (which derives from the zonal momentum equation) with the transition to non-real values of the gradient-balanced wind (which derives from the continuity, hydrostatic, and meridional momentum equations).  In contrast, the \(\urce\) value corresponding to the \(M\rce=\Omega a^2\) condition does not coincide with any mathematically unique property of \(\urce\).

\section{Hadley cell extent and ascent branch location}
\label{sec:extent}
Supposing that one or more of the conditions at Hide's theorem is met somewhere, over what latitudes does the resulting overturning circulation extend, and where does it ascend?  This section considers a series of arguments yielding progressively farther poleward predictions: those stemming directly from Hide's theorem, from the AMC model, and from the equal-area model.  We consider each for general \(\hat\theta\rce\) profiles and as applied to the canonical \(\hat\theta\rce\) profiles of LH88,
\begin{equation}
  \label{eq:lh88-forcing}
  \dfrac{\hat\theta_\mr{rce,lh88}}{\theta_0}=1+\dfrac{\deltah}{3}\left[1-3(\sinlat-\sin\maxlat)^2\right],
\end{equation}
where \(\deltah\) is an imposed fractional equator-to-pole temperature contrast, and the forcing maximizes at the latitude \(\maxlat\).  Using \eqref{eq:lh88-forcing} in \eqref{eq:u-rce}, the corresponding \(\urce\) fields are
\begin{equation}
  \label{eq:u-rce-lh88}
  u_\mr{rce,lh88}=\Omega a\coslat\left[\sqrt{1+2R\left(1-\frac{\sin\maxlat}{\sinlat}\right)}-1\right],
\end{equation}
where
\begin{equation*}
  R=\dfrac{gH\Delta_\mr{h}}{\Omega^2a^2}
\end{equation*}
is the thermal Rossby number \citep{fang_simple_1996,adam_global_2009}.  For \(\maxlat=0\), \eqref{eq:lh88-forcing} and \eqref{eq:u-rce-lh88} reduce to the original expressions of HH80.

\subsection{Constraints from Hide's theorem}
At the very least, the Hadley circulation must span all latitudes satisfying one or more of the conditions of Hide's theorem.  But this lower bound in not always especially useful, for example in the PH92 case. C.f. Figure~\ref{fig:ph92-schematic}, forcing is supercritical only within \(\sim\)18-26\degr{}N, whereas the corresponding PH92 simulation features a Hadley circulation spanning \(\sim\)25\degr{}S-30\degr{}N (\cf/ their Fig.~5a; their forcing is slightly stronger, yielding a slightly larger supercritical region of \(\sim\)12-29\degr{}N).

\begin{figure}[t]
  \centering\noindent
  \includegraphics[width=19pc,angle=0, trim={0 0.1cm 0 0.5cm}, clip]{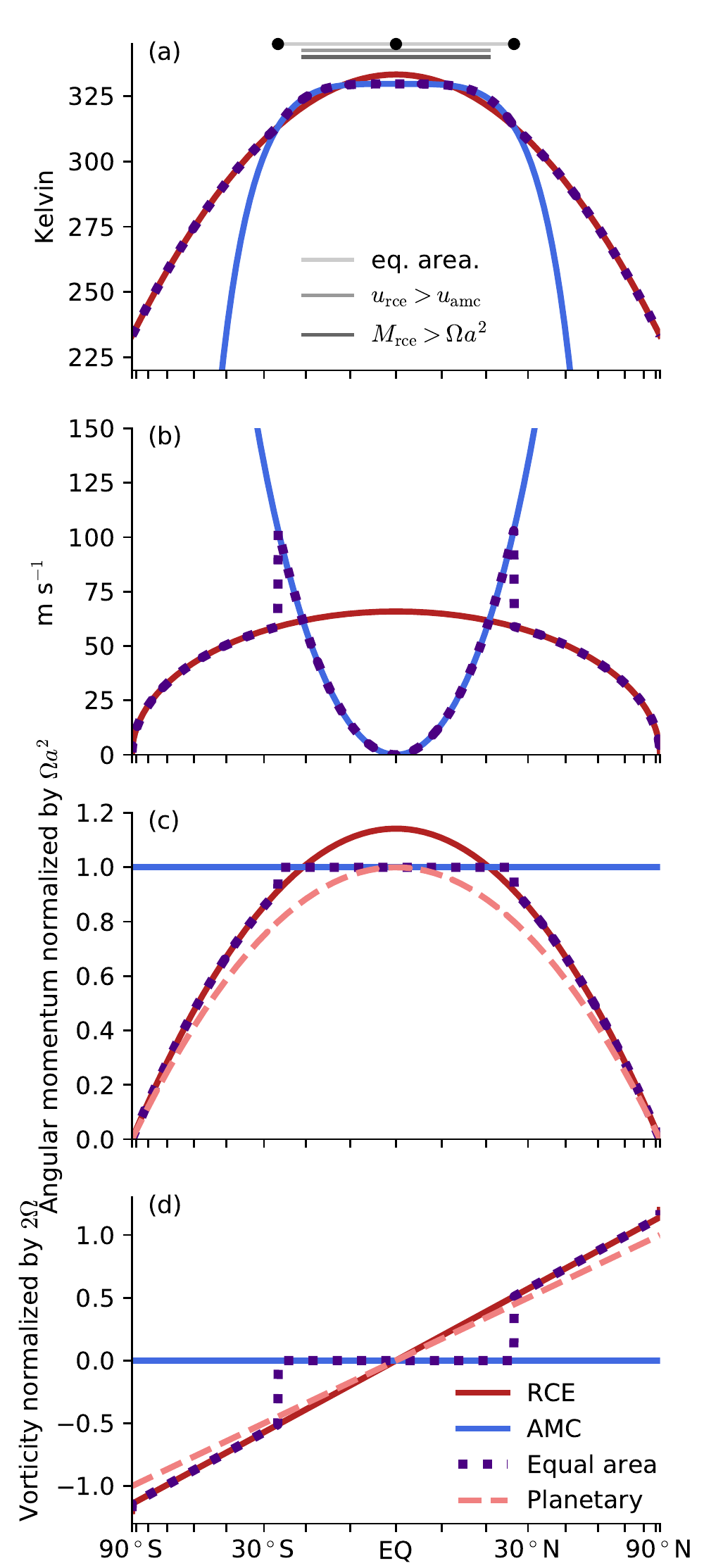}\\
  \caption{Values of (a) column-averaged potential temperature (in~Kelvin), (b) zonal wind (in \ms/), (c) absolute angular momentum normalized by \(\Omega a^2\), and (d) absolute vorticity normalized by \(2\Omega\), each corresponding to (solid red) the RCE state, (solid blue) the AMC solution, (dotted purple) the equal-area solution, and (c and d only, dashed pink) the planetary value, as a function of latitude (horizontal axis, with \(\sinlat\)-spacing), where the RCE forcing is given by \eqref{eq:lh88-forcing} with \(\maxlat=0\).  Horizontal lines at the top of (a) signify Hadley cell extent markers according to the legend in (a), with the three black dots corresponding to the cell edges of the equal-area solution.}
  \label{fig:schematic-hh80}
\end{figure}

Under the HH80 forcing, neither the \(M\rce<0\) nor \(f\etarce<0\) criteria are met at any latitude.  This is demonstrated by the solid red curves in Figure~\ref{fig:schematic-hh80}, which show (a) \(\hatthetarce\), (b) \(\urce\), (c) \(M\rce\), and (d) \(\etarce\) for \eqref{eq:lh88-forcing} with \(\maxlat=0^\circ\), with Earth's values of \(g\), \(\Omega\), and \(a\), \(H=10\)~km, and \(\Delta_\mr{h}=1/3\), yielding \(R\approx0.15\) [the other plotted elements will be discussed further below].  \(\urce\) and \(M\rce\) are maximal at the equator and decrease monotonically toward zero at either pole; the resulting meridional shear makes \(f\etarce\) \emph{more} positive than it would be in a resting atmosphere [shown as the pink dashed curve in panel (d)].\footnote{This in fact holds for the more general forcing of \(\hat{\theta}\rce/\theta_0=c_1+c_2\cos^n\lat\), where \(c_1\) and \(c_2\) are constants and \(n\geq2\) is a positive integer [of which \eqref{eq:lh88-forcing} with \(\maxlat=0^\circ\) is a special case with \(c_1=1-2\deltah/3\), \(c_2=\deltah\), and \(n=2\)].  Applying \eqref{eq:ph92-crit} to this yields
\begin{equation*}
  C_{\mr{ph92},\cos^n\lat}=-\sinlat\cos^3\lat\left[4\Omega^2a^2+n(n+2)c_2gH\cos^{n-2}\lat\right].
\end{equation*}
All of the terms within the square brackets are positive, which combined with the leading \({-}\sinlat\) term corresponds to \(f\etarce\geq0\) at all latitudes.}

\begin{figure}[t]
  \centering\noindent
  \includegraphics[width=19pc,angle=0, trim={0 0.1cm 0 0.5cm}, clip]{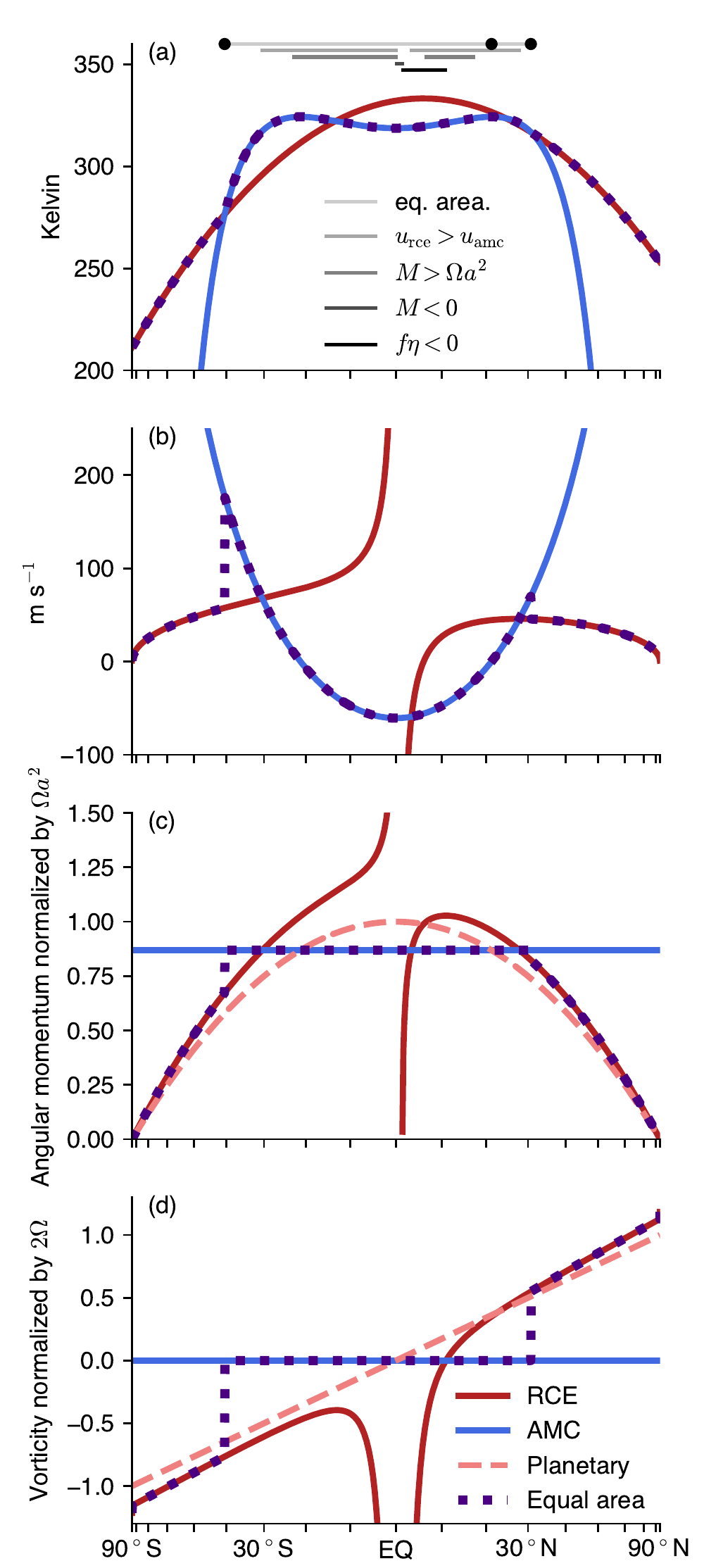}\\
  \caption{Same as Fig.~\ref{fig:schematic-hh80}, but with the RCE forcing given by \eqref{eq:lh88-forcing} with \(\maxlat=6^\circ\), and with two additional cell extent metrics in (a) as noted in the legend (neither was met at any latitude in the \(\maxlat=0^\circ\) case).  Note that, except for (d), the vertical axis spans differ from the corresponding ones of Fig.~\ref{fig:schematic-hh80}.}
  \label{fig:schematic-lh88}
\end{figure}

For the LH88 forcing, Figure~\ref{fig:schematic-lh88} repeats Figure~\ref{fig:schematic-hh80} but with \(\maxlat=6^\circ\).  Westerlies are sufficiently strong from the winter subtropics to the equator and in the \(\sim\)10\degr{}-wide span just poleward of \(\maxlat\) to generate \(M\rce>\Omega a^2\) (unlike the HH80 case, this span is not identical to that from the AMC criteria, as discussed below).  Moving across the equator toward the summer pole, \(\pdsl{\hat\theta\rce}{\lat}>0\) causes \(\urce\) to flip to non-real values, though only very near the equator.  But the meridional shear is sufficiently large poleward thereof to generate an \(f\etarce<0\) region spanning another \(\sim\)10\degr{}.  These spans are indicated by the three lower-most horizontal lines at the top of panel (a) as indicated by the legend (likewise for Figure~\ref{fig:schematic-hh80} but with the unsatisfied \(M\rce<0\) and \(f\etarce<0\) conditions omitted).  A formal treatment of the \(\eta=0\) transition under LH88 forcing is provided in Appendix B.

\begin{figure}[t]
  \centering\noindent
  \includegraphics[width=19pc,angle=0,trim={0 0 0 0}, clip]{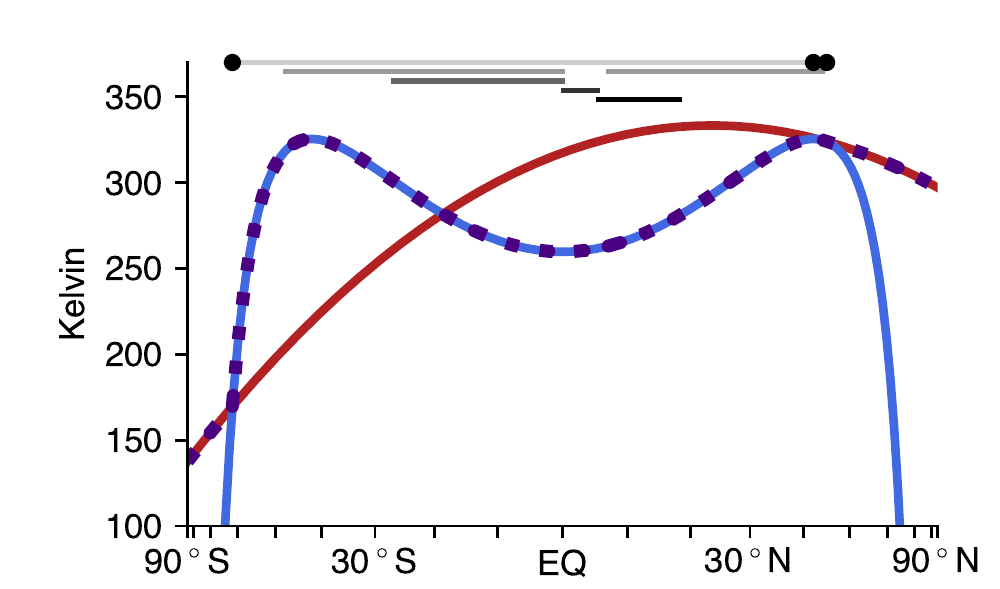}\\
  \caption{Same as Fig.~\ref{fig:schematic-lh88}(a), but with \(\maxlat=23.5^\circ\).}
  \label{fig:schematic-lh88-maxlat23}
\end{figure}

Figure~\ref{fig:schematic-lh88-maxlat23} repeats Figure~\ref{fig:schematic-lh88}(a) but with \(\maxlat=23.5^\circ\).  Now, the \(\etarce=0\) transition in the summer hemisphere occurs at 18.0\degr{}, equatorward of \(\maxlat\).  As \(\maxlat\) is moved farther poleward, \(\etarce=0\) becomes increasingly equatorward of \(\maxlat\); for example, for \(\maxlat=30^\circ\), the \(\etarce=0\) point occurs at \(\sim\)20\degr{} (not shown).

\subsection{Constraints from the AMC Hadley cell model}
\subsubsection{Conceptual basis for the AMC Hadley cell model}
The AMC model for the Hadley cells (HH80, LH88) assumes that ascent out of the boundary layer occurs at a single latitude, \(\ascentlat\), where \(u\approx0\) and hence \({M(\lat{=}\ascentlat)=\Omega a^2\cos^2\ascentlat}\).  Because viscosity is weak in the free troposphere, this angular momentum value is then homogenized over the circulation's whole expanse, requiring the AMC zonal wind profile, \cf/ LH88,
\begin{equation}
  \label{eq:u-amc}
u_\mr{amc}=\Omega a\coslat\left(\frac{\cos^2\ascentlat}{\cos^2\lat}-1\right).
\end{equation}
The blue curves in Figures~\ref{fig:schematic-hh80}(b) and \ref{fig:schematic-lh88}(b) show \(\uamc\) for \(\ascentlat=0^\circ\) and \(\ascentlat=21.2^\circ\)N, respectively (with \(\ascentlat\) determined using the equal-area model as discussed below).  For any \(\ascentlat\), \(\uamc\) vanishes at \(\ascentlat\), is mirror symmetric about the equator, and increases monotonically toward either pole from a minimum value at the equator, approaching \({+\infty}\) at the poles unless \(\ascentlat\) itself is one of the poles.  For \(\ascentlat\neq0\), \(\uamc\) is also zero at \(-\ascentlat\), easterly between \(-\ascentlat\) and \(\ascentlat\), and westerly poleward thereof.  Blue curves in Figures~\ref{fig:schematic-hh80}(c,d) and \ref{fig:schematic-lh88}(c,d) show that \(M\) is indeed constant and thus \(\eta=0\) when \(u=\uamc\).

The troposphere-averaged potential temperature field of the AMC circulation, \(\hat{\theta}\amc\), is that in gradient balance with \(\uamc\).  The \(\ascentlat=0\) case was solved by HH80 and was generalized to \(\ascentlat\neq 0\) by LH88.  The latter is
% \begin{subequations}
\begin{equation}
  \label{eq:lh88-crit-theta}
  \frac{\cm{\theta}(\lat)-\cm{\theta}_\mr{a}}{\theta_0}=-\frac{\Omega^2a^2}{2gH}\frac{(\cos^2\ascentlat-\cos^2\lat)^2}{\cos^2\lat},
\end{equation}
where \(\hat{\theta}_\mr{a}\) is the value of \(\hat{\theta}\) at \(\ascentlat\).  Blue curves in Figures~\ref{fig:schematic-hh80}(a) and \ref{fig:schematic-lh88}(a) show these for \(\ascentlat=0^\circ\) and \(\ascentlat=21.2^\circ\) (with \(\hatthetaamc\) determined from the equal-area model).  In both cases, \(\ascentlat\) corresponds to a local thermal maximum, since then \(u=0\) throughout the column as assumed in deriving \(\uamc\) and, in turn, \eqref{eq:lh88-crit-theta}.  But importantly, it does not follow that \(\ascentlat=\maxlat\).

\subsubsection{Direct \(\urce\) \vs/ \(\uamc\) comparison}
The AMC model is made globally complete by jumping at the Hadley circulation outer edges from the AMC to the RCE profile.  This is shown in the dotted purple curves of Figures~\ref{fig:schematic-hh80} and \ref{fig:schematic-lh88} (again with the cell edge latitudes determined using the equal-area model).  The jump from \(\uamc\) to \(\urce\) at the cell outer edges must occur where \(\urce\leq\uamc\), as a cell terminating where \(\urce>\uamc\) would yield an isolated local maximum in \(M\rce\) at the cell edge, thereby still violating Hide's theorem --- \ie/ if in Fig.~\ref{fig:schematic-hh80}(c) or Fig.~\ref{fig:schematic-lh88}(c) the jump from the AMC to RCE curves was upward.  It follows that the span between the two \(\urce=\uamc\) points farthest from each other constitutes a lower bound on the circulation extent (HH80).

For \(\ascentlat=0^\circ\), equating \eqref{eq:u-rce-lh88} with \eqref{eq:u-amc} yields \(\lat^*=\arccos((1+2R)^{-1/4})\), where \(\lat^*\) is the cell edge, or equivalently \(\lat^*=\arctan([(1+2R)^{1/2}-1]^{1/2})\) as expressed by HH80.  For \(\ascentlat\neq0^\circ\), \(\uamc=0\) at both \(\ascentlat\) and \(-\ascentlat\).  Provided \(\hat{\theta}\rce\) decreases monotonically from the equator to the winter pole (however modestly), then \(\urce>0\) throughout the winter hemisphere.  It follows that \(\urce>\uamc\) from the equator to some latitude poleward of \(-\ascentlat\).  Even if the meridional slope of \(\hat{\theta}\rce\) vanishes in the winter hemisphere (as it does in the simulations we present in Section~\ref{sec:crit-sims}), then \(\urce=0\), which is still more westerly than the AMC easterlies spanning from the equator to \(-\ascentlat\).  In short, an AMC cell must span at least as far into the winter hemisphere as the summer hemisphere.

\subsection{Equal-area solutions}
Using the AMC model prognostically requires a theory for \(\ascentlat\) given the RCE state.  As noted above, it is not generally the case that \(\ascentlat\approx\maxlat\).  Even for \(\maxlat=0\), off-equatorial, double ITCZs can emerge in numerical simulations such that \(\ascentlat\neq0\) \citep[\eg/][]{satoh_hadley_1994}.%  For \(\maxlat\neq0\), \(\ascentlat\) can be well equatorward of \(\maxlat\) if the thermal forcing is sufficiently weak and/or \(\maxlat\) is sufficiently poleward.  The \(M\rce>\Omega a^2\), \(M\rce<0\), \(f\etarce<0\) conditions combined provide a lower bound on the extent of the overall circulation, but in principle \(\ascentlat\) can occur anywhere within that span, constituting the boundary between summer and winter cells.  Even in the case of a single, cross-equatorial cell (for which \(\ascentlat\) necessarily is also the cell edge in the summer hemisphere), the simulations we present below indicate that \(\ascentlat\) can occur well poleward of these criteria.

The equal-area model (introduced by HH80 for \({\maxlat=\ascentlat=0}\) and extended to \({\maxlat\neq0}\), \({\ascentlat\neq0}\) by LH88) predicts \(\ascentlat\), \(\hat{\theta}\amc\), and the circulation's two poleward edges via two assumptions regarding the thermodynamic structure of the cells.  First, column-averaged potential temperature is continuous at each cell edge; second, flow within each cell conserves energy.  If \(\hat\theta\rce\) is symmetric about the equator and \(\ascentlat=0\), the two cells are mirror symmetric, yielding two equations to solve for the two unknowns.  See Equations 8-11 of LH88.

In the small angle limit, the HH80 equal-area solution terminates at \(\lat_\mr{H}=(5R/3)^{1/2}\), a factor of \((5/3)^{1/2}\) poleward of where \(\urce=\uamc\).  Without the small angle assumption, an analytical solution is no longer attainable, but the resulting expression (\Eq/~17 of HH80) is readily solvable numerically and always yields a cell terminating poleward of the \(\urce=\uamc\) line.

Purple dotted curves in Figures~\ref{fig:schematic-hh80}-~\ref{fig:schematic-lh88-maxlat23} show the equal-area solutions for the given forcing.  Also overlaid as gray horizontal lines are the spans of the various extent metrics --- from the equal-area solution (with dots denoting the three cell edges), \(\urce>\uamc\) with the \(\ascentlat\) value taken from the equal-area model, and the conditions of Hide's theorem.  In the HH80 case (Figure~\ref{fig:schematic-hh80}), because \(\ascentlat=0\), the \(M\) value being homogenized by \(\uamc\) is the equatorial value, \(M=\Omega a^2\), such that the \(M\rce>\Omega a^2\) and \(\urce>\uamc\) criteria are identical.  The \(f\etarce<0\) and \(M\rce<0\) conditions are omitted, because they do not occur at any latitude as described above.

In the LH88 case with \(\maxlat=6^\circ\) (Figure~\ref{fig:schematic-lh88}), because \(\maxlat\neq0\), \(\urce\) is no longer symmetric about the equator, and now regions of \(\Mrce<0\) and \(f\etarce<0\) emerge in the summer hemisphere.  In addition, \(\ascentlat\neq\maxlat\), such that the \(\urce>\uamc\) and \(M>\Omega a^2\) spans are no longer identical, with the former yielding a larger minimum circulation extent.  and the value of \(M\) being homogenized depends on the solution for \(\ascentlat\).  Notice that the equal-area \(\ascentlat\) prediction is appreciably poleward of \(\maxlat\) and of the \(\etarce=0\) transition.

For \(\maxlat=23.5^\circ\) (Figure~\ref{fig:schematic-lh88-maxlat23}), the equal-area model predicts a circulation spanning 61.7\degr{}S-44.8\degr{}N --- implausibly large for Earth, despite all parameters taking Earth-like values and \(\maxlat\) sitting far equatorward of where it would for a true solsticial RCE solution, namely at the summer pole.  And as \(\maxlat\) is moved farther poleward, the separation between \(\maxlat\) and the equal-area \(\ascentlat\) continues to grow.  Although the equal-area solution never truly becomes pole-to-pole at Earth's rotation rate \citep{guendelman_axisymmetric_2019}, these properties are clearly inadequate for Earth's solsticial Hadley circulation.

\subsection{Caveats}
Hide's theorem holds for any nonzero viscosity, and in atmospheres with sufficiently large \(\nu\) the appropriate model of the Hadley cells is the viscous, linear one \citep{schneider_axially_1977-1,fang_solution_1994}.  But as a model for most terrestrial planets including Earth, the nearly inviscid, angular momentum conserving model is more appropriate; see \citet{fang_solution_1994} for a formal treatment of this regime separation based on the relative values of the Ekman and Rossby numbers.  And the nonlinear, nearly inviscid solution is not equal to the linear, viscous solution in the limit as \(\nu\rightarrow0^+\) (HH80).

On Earth, baroclinic eddies modulate the Hadley cells' extent and overturning strength \citep[\eg/][]{walker_response_2005,korty_extent_2008,levine_baroclinic_2015,singh_exploring_2016,singh_eddy_2016} and decelerate the zonal wind of the annual mean Hadley cells well below the AMC limit \citep{held_large-scale_1985,walker_eddy_2006}.  Nevertheless, large expanses of cross-equatorial, zonally confined monsoons and zonal mean Hadley cells during solsticial seasons do approximately behave as in the AMC regime, with local Rossby numbers \(\mathrm{Ro}\equiv-\zeta/f\sim0.6\) compared to \({\mathrm{Ro}=1}\) for the true AMC solution and \({\mathrm{Ro}=0}\) for the purely eddy-dominated case \citep[\eg/][]{schneider_eddy-mediated_2008,bordoni_monsoons_2008,bordoni_regime_2010}.  Accordingly, axisymmetric theory \citep[in its modern ``convective quasi-equilibrium'' form appropriate for moist, convecting atmospheres, \cf/][]{emanuel_thermally_1995} remains of value in --- and is the dominant existing theoretical paradigm for --- these contexts \citep[\eg/][]{nie_observational_2010}.  The importance of baroclinic eddies diminishes as either planetary rotation rate or radius decrease, leading to a ``global Tropics'' regime once the Rossby radius of deformation exceeds the planetary scale \citep{Williamsrangeunityplanetary1982,mitchell_dynamics_2006,faulk_effects_2017}.  In addition to the solar system's slow rotators of Venus and Titan, in all likelihood this characterizes many ``habitable zone'' exoplanets identified already or likely to be identifiable by existing and planned telescope missions, due to orbital dynamical constraints \citep[\eg/][]{showman_atmospheric_2014}.

Even in axisymmetric atmospheres, multiple processes prevent the Hadley cells from ever actually reaching the AMC limit, as has been noted by many authors (\eg/ HH80, LH88).  Ascent always occurs over some finite latitudinal width, leading to different streamlines leaving the boundary layer with different \(M\) values.  Within the ascending branch, convective momentum mixing may be non-negligible \citep{schneider_axially_1977-1,schneider_axially_1977}, although typically weak vertical shear there makes it not obviously a large term \citep[\cf/][]{held_large-scale_1985,zheng_response_1998}.  Once beyond the ascending branch, nonzero viscosity, however small, then generates mixing across tightly packed horizontal streamlines in the cell's upper branch where \(\pdsl{u}{z}\) is large and in the descending branch.  \citet{fang_simple_1996} derive an analytic ``viscous correction'' to their otherwise inviscid solution that accounts for this, which acts to smear out the otherwise step changes in temperature and zonal velocity at their cell edges (see their Figure~4).

Separately, motionless air advected into the upper branch retards the upper level flow, and this has been compactly addressed in the ``1 1/2'' layer shallow water models of \citet{shell_abrupt_2004} and \citet{adam_global_2009}.  These yield an equinoctial Hadley cell with nearly-uniform height in the ascending branch connected to an AMC subsiding branch, and arguably they more accurately capture the momentum structure of the HH80 numerical solutions than does the standard AMC solution.  Moreover, the circulation in this 1 1/2 layer model spans farther into the winter hemisphere than the summer hemisphere \citep{adam_global_2010}, as in the unmodified AMC model as described above.

% As noted by \cite{fang_simple_1996}, \(\uamc\) does not vary with height, which by gradient balance requires that cell meridional equivalent potential temperature gradients are exactly zero in the free troposphere.  As such, the meridional variations in \(\hat\theta\amc\) must arise entirely within the boundary layer.  Thus, the true AMC solution could be discontinuous in potential temperature at the cell edges at individual levels while being continuous in the column average.

% As noted by \cite{fang_simple_1996} and \cite{singh_exploring_2016}, if the truly free troposphere is truly inviscid and the PBL is strongly damped, then streamlines must exit and enter the PBL at the same latitude.  Their angular momentum value is set by the planetary angular momentum where they exit the PBL, this value remains when they re-enter the PBL, and \(u\approx0\) within the PBL, such that unless \(M\) is discontinuous across the PBL top, the parcel must return at the same latitude.  Although surface winds are not actually zero and the free troposphere is not truly inviscid, it is interesting to note that angular momentum contours and streamlines in our simulations are remarkably symmetric about the equator, exiting the PBL in the summer hemisphere and re-entering it in the winter hemisphere at nearly the same value.

\section{Results from sub- and supercritically forced simulations}
\label{sec:crit-sims}
We have argued that the span of supercritical forcing is a lower bound on the cross-equatorial cell extent in the summer hemisphere, and that cross-equatorial, AMC cells are bound to be relatively symmetric in extent about the equator.  This section presents evidence for those claims via simulations in an idealized GCM in which \(\etarce=0\) is made to occur either near \(\maxlat\) or well equatorward thereof.

\subsection{Description of the idealized dry GCM}
We use the dry idealized GCM of \citet{schneider_tropopause_2004}, which solves the primitive equations on the sphere with no topography using a spectral dynamical core.\footnote{As noted by \citet{adam_global_2009}, spectral solvers are not ideal for cases such as the axisymmetric Hadley cells in which the solutions are expected to have discontinuities.}  All parameters take Earth-like values except as otherwise noted.  The simulations are axisymmetric by way of exactly axisymmetric initial conditions and boundary conditions.

Convective adjustment relaxes temperatures in statically unstable columns toward the dry adiabatic lapse rate, \(\Gamma_\mr{d}\equiv g/c_p\), as would hold in a state of dry RCE, over a globally uniform 4~day timescale.  Radiative transfer is approximated by Newtonian cooling, wherein temperatures are relaxed toward a prescribed field (described in the next subsection) over a timescale that is 50~days throughout the free troposphere and decreases linearly in \(\sigma\) from that value at the PBL top to 7~days at the surface.  The treatment of dissipative processes is standard, but given the potential nuances relating to the nearly inviscid assumptions, is described in full in Appendix C.

\subsection{Imposed RCE temperature profiles}
The equilibrium temperature fields at the lowest model level (which we refer to as the surface values) are based on \eqref{eq:lh88-crit-theta}, modified as described below, with a dry adiabatic lapse rate from the surface to a specified tropopause temperature.  The atmosphere is isothermal from the tropopause upwards.  The dry adiabatic stratification, along with the dry adiabatic convective adjustment, ensures little distinction between the imposed equilibrium temperatures and a true RCE solution.  This is confirmed via computing RCE solutions for a subset of our cases by repeating them with all advective terms suppressed (not shown); the temperature structure is always almost exactly dry adiabatic from the surface to the tropopause.

Because our model is not Boussinesq, rather than using \eqref{eq:lh88-crit-theta} we use the analogous expression derived by E95 appropriate for (dry or moist) atmospheres obeying convective quasi-equilibrium \citep[CQE][]{emanuel_large-scale_1994}:
\begin{equation}
  \label{eq:e95-crit-theta}
  \thetab=\theta_\mr{ba}\exp\left[-\frac{\Omega^2a^2}{2c_p(T_\mr{s}-T_\mr{t})}\frac{(\cos^2\ascentlat-\cos^2\lat)^2}{\cos^2\lat}\right],
\end{equation}
where \(\thetab\) is boundary layer potential temperature (replaced in moist atmospheres by \(\theta_\mr{eb}\), the subcloud equivalent potential temperature), \(\theta_\mr{ba}\) is the value of \(\thetab\) at \(\ascentlat\), \(\Tsfc\) is the surface temperature, \(\Ttrop\) is the temperature at the tropopause, and \(\Tsfc-\Ttrop\), has been assumed constant.  Although \eqref{eq:e95-crit-theta} is an expression for the dynamically equilibrated AMC state, here we use it (modified as described below) as the RCE state to which the model is being relaxed toward.  Therefore, in the context of these simulations only, it should be interpreted as a forcing, with \(\ascentlat\) replaced by \(\maxlat\); the simulations will generate their own \(\ascentlat\) that in general need not be the same as the imposed \(\maxlat\).

To generate cells that extend either to the vicinity of \(\maxlat\) or well equatorward thereof, we insert a multiplicative factor, \(\alpha\), into the exponential, set to 0.5 or 2.0, respectively.  Temperatures are uniform beginning 10\degr{} past the forcing maximum \(\maxlat\), since otherwise except for \(\maxlat=90^\circ\) the profile would drop toward absolute zero.  To break hemispheric symmetry, at 10\degr{}N we switch from the original profile to its local tangent, following this to 10\degr{}S, and then setting temperatures uniformly to their value at 10\degr{}S farther south.

\subsection{Simulations performed}
\begin{figure*}[t]
  \centering\noindent
  \includegraphics[width=39pc, angle=0, trim={0cm 0cm 0cm 0cm}, clip]{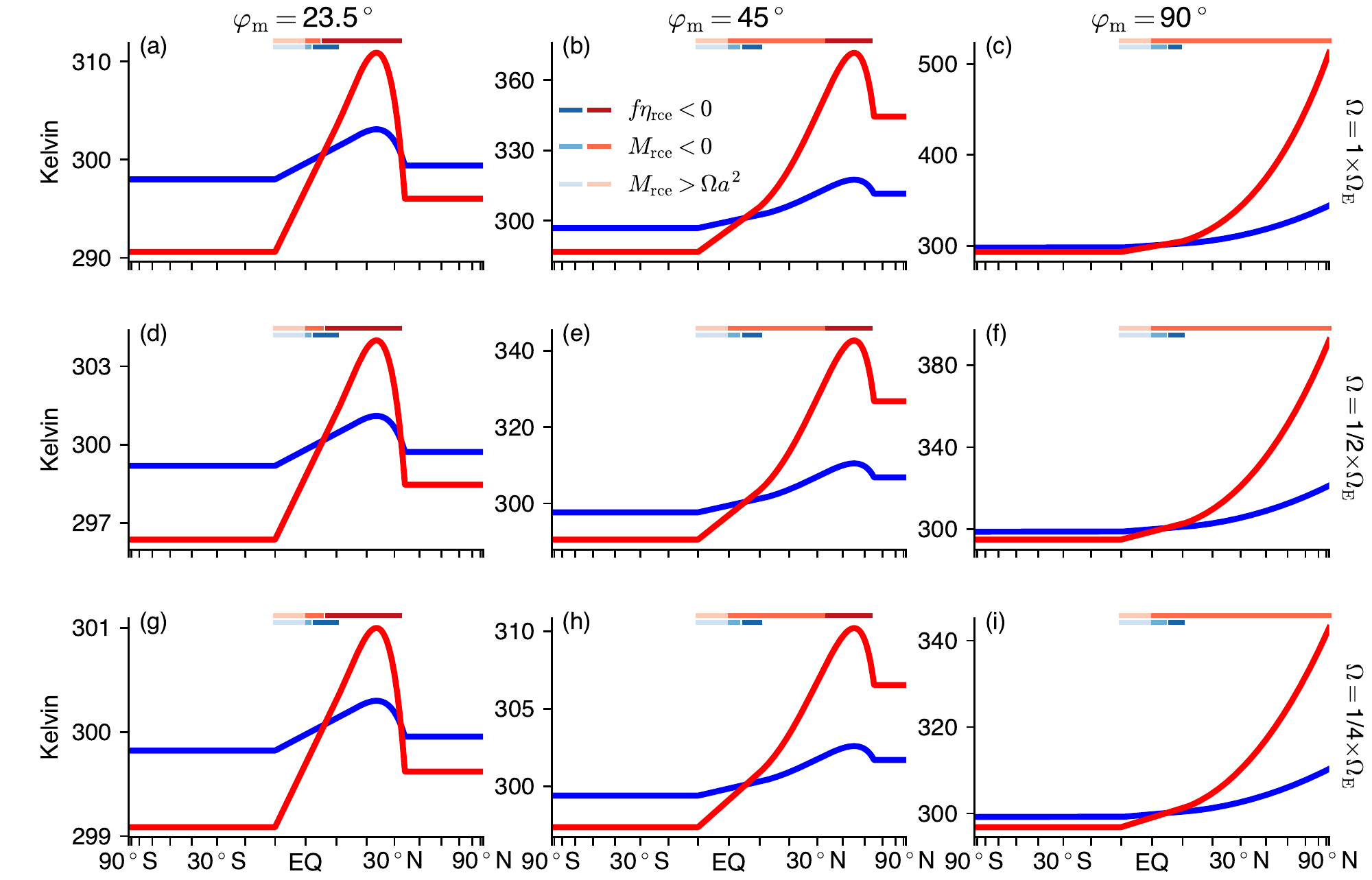}\\
  \caption{Equilibrium surface temperature distribution (in~Kelvin) that is relaxed toward in each simulation: (left to right) forcing maximum latitude \(\latmax\), (top to bottom) planetary rotation rate, and (thick curves) value of \(\alpha\) (blue for \(\alpha=0.5\), red for \(\alpha=2.0\)).  Horizontal lines at the top of each panel correspond to the extent metrics as indicated in the legend in panel, with blue shades for the \(\alpha=0.5\) case and red shades for \(\alpha=2.0\).  Note different vertical axis spans in each panel.}
  \label{fig:sims-forcings}
\end{figure*}

We perform one \(\alpha=0.5\) and one \(\alpha=2.0\) simulation for each of \(\maxlat=\)23.5\degr{}, 45\degr{}, and 90\degr{} and for each of \(\Omega=1\times\), \(1/2\times\), and \(1/4\times\Omega_\mr{E}\), where \(\Omega_\mr{E}\) is Earth's rotation rate.  Table~\ref{tab:sub-sup-sim-params} lists the forcing parameters for each simulation, and Figure~\ref{fig:sims-forcings} shows the imposed equilibrium surface temperature profiles, as well as the spans where the \(f\etarce<0\), \(M\rce<0\), and \(M\rce>\Omega a^2\) conditions for cell extent are met.  The linear increase in temperature spanning the equator in all cases ensures \(M\rce>\Omega a^2\) on the winter-side of the equator and \(M\rce<0\) on the summer-side.  In the \(\alpha=2.0\), \(\maxlat=90^\circ\) cases, the \(\Mrce<0\) region extends all the way to the pole.  In all other cases, poleward of the \(M\rce<0\) region there exists a finite region where \(f\etarce<0\) in the summer hemisphere.  The combined range of these extent conditions spans \({\sim}10^\circ\)S-10\degr{}N in the  \(\alpha=0.5\) cases (\ie/ those latitudes with the linear temperature profile), compared to \(\maxlat\) or somewhat poleward thereof in the \(\alpha=2.0\) cases.

Parameter values were chosen on an ad hoc basis for each \((\Omega,\maxlat)\) pair in order to generate supercritical forcing from the equator to \(\maxlat\) in the \(\alpha=2.0\) case with the minimal meridional temperature variation possible, and with surface temperatures near the equator \(\sim\)300~K in all cases (so that the tropospheric depth is similar across simulations, at least away from \(\maxlat\)).  For forcing maxima well removed from the Tropics at Earth's rotation rate, it is difficult to generate \(f\etarce<0\) near \(\maxlat\) while keeping a realistic tropopause depth.  For this reason, the tropopause is set to 100~K in our forcing profiles in some of the \(1\times\Omega_\mr{E}\) cases, in which the forcing surface temperature at the maximum can exceed 500~K.  Under dry adiabatic stratification, this yields an effective tropopause height of \(\sim\)40~km.  Stated another way, if we require that the tropospheric depth remains roughly Earth-like, we are unable to violate Hide's theorem near \(\maxlat\) at Earth's rotation rate or faster for high-latitude \(\maxlat\).

Table~\ref{tab:sub-sup-sim-params} also lists diagnosed values of \(\Delta_\mr{h}\), computed as the global maximum minus the global minimum of the surface forcing temperature divided by its global mean, and the corresponding diagnosed value of \(R\).  These were diagnosed after the simulations were complete, \ie/ they were not tuned for, so it is interesting that the \(R\) values are quite similar across \(\Omega\) values, particularly for \(\alpha=2.0\).  For example, \({R\approx.1{-}1.2}\) in all three \(\maxlat=90^\circ\), \(\alpha=2.0\) cases.  Also note that these \(R\) values are much less than the critical values required for the \(f\etarce<0\) condition to be satisfied in the case of LH88 forcing shown in Figure~\ref{fig:eta-lh88-phase-diagram}.  This is because the \(\eta\) distribution depends not just on the total magnitude of meridional temperature variations but on the \emph{shape} of those variations --- as discussed previously, the \(\hat\theta\rce\) profiles in Figure~\ref{fig:sims-forcings} ``ramp up'' with positive meridional curvature from the equator nearly to \(\maxlat\), rather than the LH88 cases [Figure~\ref{fig:schematic-lh88}(a)] in which the meridional curvature is negative throughout the domain.

\begin{table}[t]
\caption{Forcing parameters of each simulation performed.  From left to right: ratio of planetary rotation rate to Earth's rotation rate, where \(\Omega_\mr{E}\) is Earth's rotation rate (s\inv{}); latitude of forcing maximum (\degr); value of \(\alpha\) (dimensionless); tropopause temperature (K); difference between surface and tropopause temperatures used in the expression to generate the forcing (K); temperature at the forcing maximum (K); largest fractional temperature variation (dimensionless); and diagnosed thermal Rossby number (dimensionless).  Parameter values were chosen so that \(\thetab\sim300\)~K at the equator in all cases.}\label{tab:sub-sup-sim-params}
\begin{center}
\begin{tabular}{lrrrrrrr}
\hline\hline
\(\Omega/\Omearth\) & \(\maxlat\) & \(\alpha\) & \(\Ttrop\) & \(T\sfc-\Ttrop\) & T\(_{\text{max}}\) & \(\Delta_\mr{h}\) & \(R\)\\
\hline
1 & 23.5 & 0.5 &     &     & 303.1 & 0.01 & 0.01 \\
                    &      & 2.0 &     &     & 310.9 & 0.07 & 0.05 \\
                    & 45   & 0.5 & 100 & 250 & 317.5 & 0.07 & 0.08 \\
                    &      & 2.0 &     &     & 371.6 & 0.27 & 0.32 \\
                    & 90   & 0.5 &     & 400 & 344.1 & 0.15 & 0.67 \\
                    &      & 2.0 &     &     & 512.2 & 0.28 & 1.25 \\
\hline
0.5 & 23.5 & 0.5 &     &     & 301.1 & 6\(\times10^{-3}\) & 0.01 \\
                      &      & 2.0 &     &     & 304.0 & 0.03 & 0.05 \\
                      & 45   & 0.5 &     &     & 310.5 & 0.04 & 0.08 \\
                      &      & 2.0 &     &     & 342.7 & 0.17 & 0.32 \\
                      & 90   & 0.5 &     & 200 & 321.2 & 0.07 & 0.28 \\
                      &      & 2.0 &     &     & 392.1 & 0.31 & 1.16 \\
\hline
0.25 & 23.5 & 0.5 &     &     & 300.3 & 2\(\times10^{-3}\) & 0.01 \\
                       &      & 2.0 &     &     & 301.0 & 6\(\times10^{-3}\) & 0.05 \\
                       & 45   & 0.5 &     &     & 302.6 & 0.01 & 0.08 \\
                       &      & 2.0 &     &     & 310.2 & 0.04 & 0.32 \\
                       & 90   & 0.5 &     &     & 310.3 & 0.04 & 0.28 \\
                       &      & 2.0 &     &     & 342.9 & 0.15 & 1.13 \\
\hline
\end{tabular}
\end{center}
\end{table}

All simulations are run for 1200~days starting from an isothermal, resting state, with results presented as averages over the last 1000.  Though a statistically steady state is achieved throughout the domain generally within 100~days, regular transient symmetric instabilities persist throughout the integration over much of the extent of the Hadley cells (not shown).  As one example, in the simulation at Earth's rotation rate with \(\maxlat=45^\circ\) and \(\alpha=2.0\), equatorward propagating features are prominent from roughly 35\degr{}S to 10\degr{}S.

\subsection{Results}
\begin{figure*}[t]
  \centering\noindent
  \includegraphics[width=39pc, angle=0, trim={0 0 0 0}, clip]{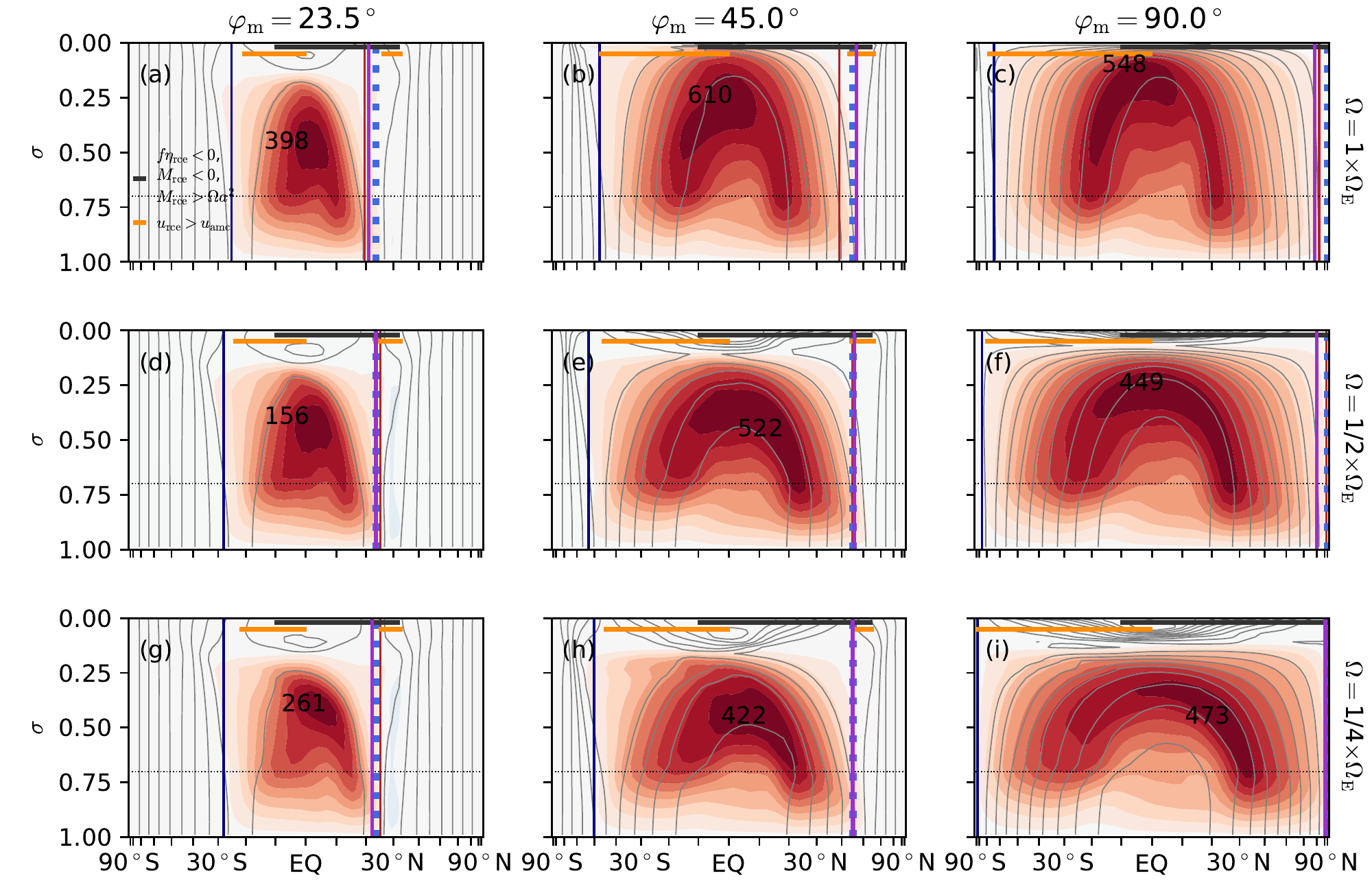}\\
  \caption{(Filled contours) meridional overturning streamfunction and (grey contours) absolute angular momentum fields in the simulations with \(\alpha=2.0\) (see text for explanation of the experimental setup) with (left to right) the forcing maximum at 23.5, 45, and 90\degr{}N, and (top to bottom) planetary rotation rate 1, 1/2, and 1/4\(\times\)Earth's rotation rate.  In each panel, the contour interval for the streamfunction is 10\% of the value at the cell center, labeled by the red star and adjacent text in 10$^{9}$ kg s$^{-1}$, with red shades denoting positive values and blue shades negative values, and the contour interval for the angular momentum is 10\% of the planetary angular momentum at the equator.  The red vertical lines denote the cross-equatorial Hadley cell's edges in the (solid) winter and (dashed) summer hemisphere, based on where the streamfunction reduces to 10\% of its maximum at the same level.  The purple solid line denotes the effective \(\ascentlat\), computed as described in the text.  The blue dotted lines correspond to the location of the forcing maximum \(\maxlat\).  The dotted horizontal line marks the planetary boundary layer top of \(\sigma=0.7\).}
  \label{fig:supercrit-streamfunc}
\end{figure*}

\begin{figure*}[t]
  \centering\noindent
  \includegraphics[width=39pc, angle=0, trim={0 0 0 0}, clip]{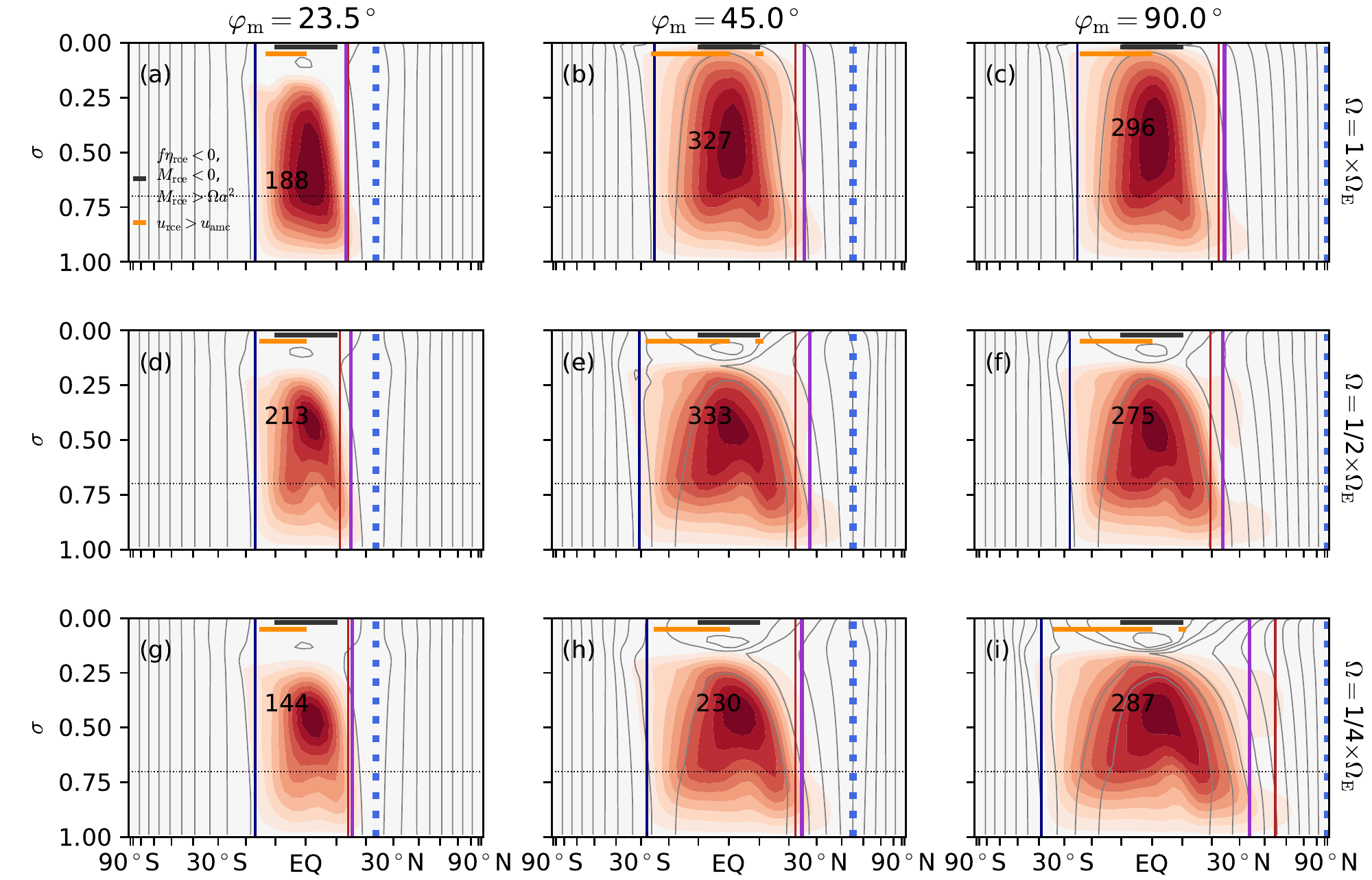}\\
  \caption{As in Figure~\ref{fig:subcrit-streamfunc}, but for the \(\alpha=0.5\) simulations.}
  \label{fig:subcrit-streamfunc}
\end{figure*}

Figure~\ref{fig:supercrit-streamfunc} shows the meridional overturning streamfunctions in each \(\alpha=2.0\) case, and Figure~\ref{fig:subcrit-streamfunc} shows the same for the \(\alpha=0.5\) cases, with angular momentum contours, the computed Hadley cell edge latitudes, and \(\maxlat\) overlaid.  The cell edges are computed using the standard metric of where the streamfunction at the sigma level of its maximum decreases to 10\% of that maximum \citep{walker_eddy_2006}, but weighted as in \cite{singh_limits_2019} to account for the weakening influence on the streamfunction of meridians converging toward the pole.  Symbolically,
\begin{equation}
  \label{eq:edge-def}
  \dfrac{\Psi(\lat_\mr{h},\sigma_\mr{max})}{\cos\lat_\mr{h}}=  0.1\dfrac{\Psi_\mr{max}}{\cos\lat_\mr{max}},
\end{equation}
where \(\Psi\) is the Eulerian-mean streamfunction, \(\lat_\mr{h}\) is the cell edge, and \(\Psi_\mr{max}\) is the streamfunction global maximum magnitude, which occurs at latitude \(\lat_\mr{max}\) and sigma level \(\sigma_\mr{max}\).

A few features are consistent across all simulations.  First, the nonzero forcing gradient spanning the equator necessitates an overturning cell in all cases.  Second, the streamlines of those cells are nearly coincident in the free troposphere with angular momentum contours.  Third, the Hadley circulations all comprise a single cross-equatorial cell, with no discernible summer cell.  And, fourth, the cross-equatorial cells extend to nearly the same latitudes in the winter and summer hemispheres, typically somewhat farther into the winter hemisphere.

Notable in the \(\alpha=2.0\) cases are very strong equatorial jumps as described by \cite{pauluis_boundary_2004}, with some streamlines bending up out of the boundary layer near \(\pm10^\circ\) and rising by as much as \({\sim}0.4\) in \(\sigma\).  This jumping does not appear to affect the features of our interest; simulations in which the jump is suppressed via a stronger cross-equatorial temperature gradient do not differ qualitatively outside the deep tropics (not shown).

Also overlaid are the ranges where any of the \(M\rce>\Omega a^2\), \(M\rce<0\), and \(f\etarce<0\) conditions for cell extent are met, and separately where the \(\urce>\uamc\) condition is met.  The value of \(\ascentlat\) used to compute \(\uamc\) is diagnosed from the minimum value of angular momentum at the equator in the simulations, \(M_\mr{eq,min}\).  Given that value, then \(\ascentlat\approx\arccos(\sqrt{M_\mr{eq,min}/\Omega a^2})\).  Roughly speaking, this amounts to finding the angular momentum contour coincident with the circulation's topmost streamline at the equator and tracing it back to the surface in the summer hemisphere \citep{singh_limits_2019}.  This value (shown as a vertical purple line) is generally close to the edge diagnosed from the streamfunction and for the \(\alpha=2.0\) cases is near \(\maxlat\) also.

This rough coincidence of \(\ascentlat\) and \(\maxlat\) for \(\alpha=2.0\) leads to a single, nearly pole-to-pole cell in the polar-maximal forcing cases.  The biggest offsets of \(\ascentlat\) and the cell edge from \(\maxlat\) occur in the \(\Omega=1\times\Omega_\mr{E}\) and \(\maxlat=90^\circ\), for which the computed cell edge metric sits at \(\sim\)69\degr{}.  However, the streamfunction retains its sign all the way to the pole, and the cell is nevertheless global in scale (zonal winds near the equator approach 400~\ms/ in the stratosphere in this case; not shown).  This is a remarkable contrast to Earth's present-day cross-equatorial Hadley cell, whose summer hemisphere edge never extends beyond \(\sim\)15\degr{} in either hemisphere \citep[based on the zero crossings of the overturning streamfunction shown in Figure~4 of][]{adam_seasonal_2016}.

For the \(\alpha=0.5\) cases, the cells typically terminate well equatorward of the forcing maximum --- in the most extreme case, with \(\Omega=1/2\times\Omega_\mr{E}\) and \(\maxlat=90^\circ\), \(\ascentlat\) and the cell edge are near 20\degr{}.  The poleward extent of supercritical forcing is typically equatorward of the cell edge and \(\ascentlat\) by \(\gtrsim10^\circ\), but nevertheless across all \(\alpha=0.5\) simulations is a more accurate predictor of cell extent than is \(\maxlat\).

\begin{figure*}[t]
  \centering\noindent
  \includegraphics[width=39pc, angle=0, trim={0cm 0cm 0cm 0cm}, clip]{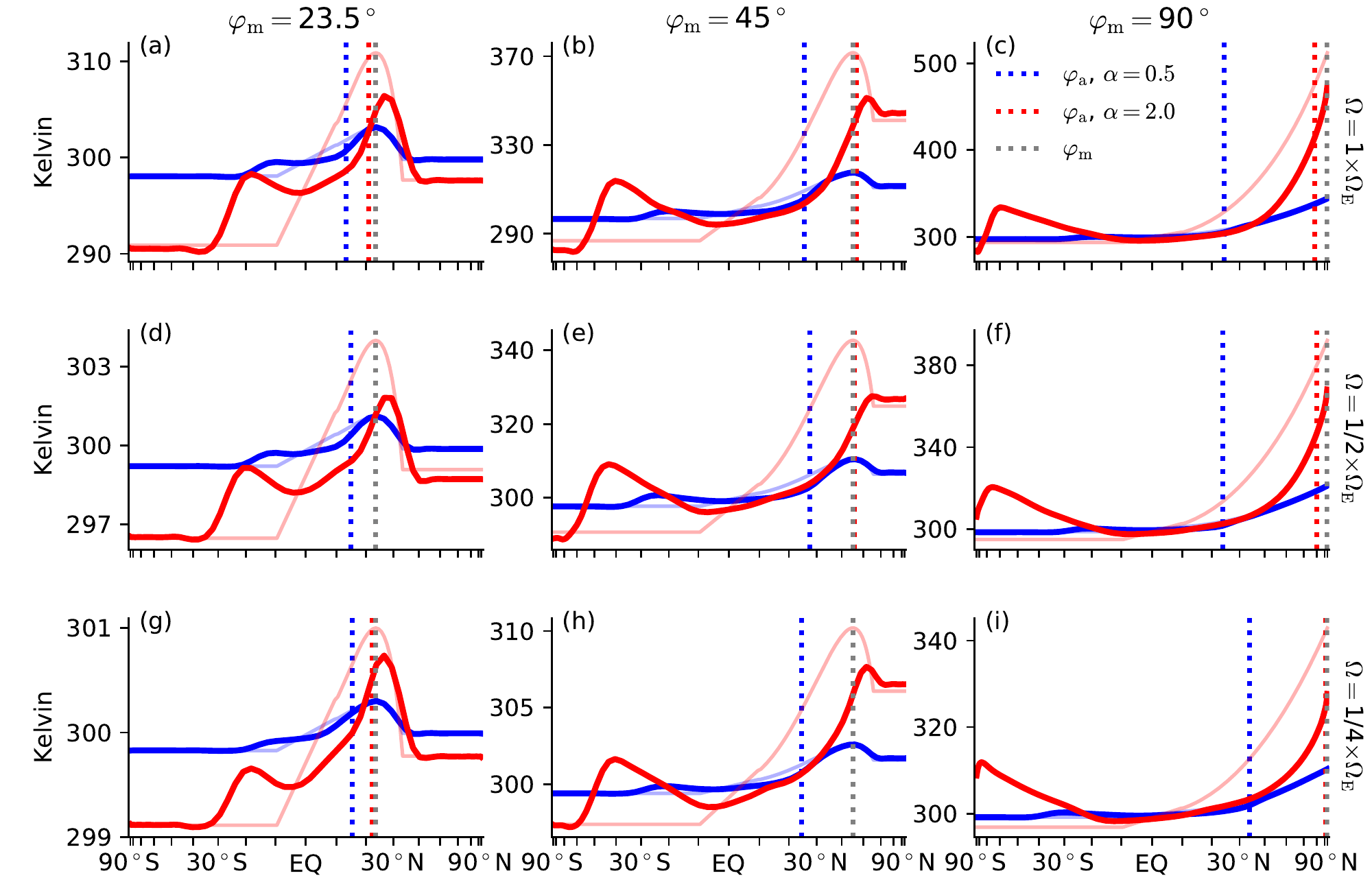}\\
  \caption{(Thick solid curves) potential temperature (in~K) at the \(\sigma\approx0.85\) level in the (blue) \(\alpha=0.5\) and (red) \(\alpha=2.0\) cases and (thin solid curves) the corresponding forcing values, with panels oriented as in Figure~\ref{fig:sims-forcings}.  As indicated in the legend in panel (a), vertical dotted lines are the effective ascent latitude for (blue) \(\alpha=0.5\) and (red) \(\alpha=2.0\) and (gray) the forcing maximum latitude.  Note different vertical axis spans in each panel.}
  \label{fig:temp-sfc-sims}
\end{figure*}

Figure~\ref{fig:temp-sfc-sims} shows \(\thetab\), which we diagnose as the value of \(\theta\) at \(\sigma\approx0.85\) in all cases.  The \(\alpha=0.5\) cell terminates near where \(\pdsl{\thetab}{\lat}\approx0\), but sits just poleward of an inflection point rather than just equatorward of a maximum as suggested by \citet{prive_monsoon_2007}.  The flattening of \(\thetab\) equatorward of the ascent branch can be interpreted \cf/ \citet{schneider_eddy-mediated_2008} as caused by the southerly flow in the cell's lower branch advecting \(\thetab\) up-gradient.  This flattens \(\thetab\) up to where the meridional flow diminishes, at which point \(\thetab\) begins increasing sharply with latitude moving farther poleward.  Conversely, with \(\alpha=2.0\) temperatures are minimum near the equator and increase moving into either hemisphere as needed to generate the strong easterlies necessary for the AMC cell.

Because we are not using the HH80 or LH88 forcing profiles in these simulations, we cannot explicitly compare to the equal-area model predictions.  But Figure~\ref{fig:temp-sfc-sims} shows that equal-area-like behavior is occurring (compare the thick dark lines to the corresponding thin pale lines, the latter being \(\hatthetarce\)).  The dynamically equilibrated \(\hat\theta\) fields intersect the \(\hat\theta\rce\) fields at low latitudes where the former are flat and the latter are quite steep, in the winter hemisphere from below, and in the summer hemisphere from above, yielding areas between the two curves that, at least by eye, roughly cancel.  But importantly, in all simulations \(\ascentlat\leq\maxlat\), in contradiction to the equal-area prediction.  Comparison to the equal-area model is further hindered by the fact that, especially for the global-scale cells, the ascending branch is sufficiently wide that the AMC assumption of a single \(M\) value throughout the cell becomes problematic.

\section{Summary}
\label{sec:summ}
We have presented theoretical arguments and numerical modeling results pertaining to the  extent of the cross-equatorial Hadley cell under solsticial forcing, utilizing steady, dry, axisymmetric, nearly inviscid theory.  By Hide's theorem, a state of latitude-by-latitude radiative-convective equilibrium (RCE) is impossible if its distribution of absolute angular momentum (\(M\rce\)) exhibits any local extrema away from the surface, which occurs if \(M\rce<0\) or \(M\rce>\Omega a^2\) at any latitude or if the corresponding absolute vorticity (\(\etarce\)) exhibits \(f\etarce<0\) at any latitude.  A more general form of the \(f\etarce<0\) condition holds even in zonally varying and/or purely inviscid atmospheres (E95; see our Appendix A), and a Hadley circulation must span all latitudes that meet one of these conditions.  But in the winter hemisphere, typically only the \(\Mrce>\Omega a^2\) condition is met, and only very near the equator, making it Hide's theorem by itself a poor predictor of how far a cross-equatorial cell will extend into the winter hemisphere.

The angular momentum conserving (AMC) Hadley cell models, yield cells that must span all latitudes where the RCE zonal wind (\(\urce\)) exceeds the AMC zonal wind (\(\uamc\)).  This \(\urce>\uamc\) lower bound on the circulation extent provides a simple heuristic argument for why cross-equatorial cells typically extend at least as far into the winter hemisphere as into the summer hemisphere in axisymmetric atmospheres.  However, using the AMC model prognostically requires a prediction for the latitude at which ascent is concentrated (\(\ascentlat\)) and thus where the planetary angular momentum value gets imparted to the free troposphere.

The equal-area model combines the AMC assumptions with assumptions of energy conservation by the cells and continuity of potential temperature at their edges and generate predictions given the RCE state for the edges of each overturning cell, unlike the \(\urce>\uamc\) condition that sets a lower bound for the extent across all cells.  However, they can be solved analytically only in the on-equatorial forcing, small-angle case.  Much more problematic, as the latitude where the forcing maximizes (\(\maxlat\)) is moved poleward, the predicted Hadley circulation extent becomes implausibly large for Earth, even when all parameters are Earth-like.

Simulations in an idealized, dry GCM in which temperatures are relaxed at each timestep toward a specified RCE field that is either subcritical or supercritical in the summer hemisphere up to a specified \(\maxlat\) show the utility of Hide's theorem.  When the forcing is subcritical outside the deep tropics, Hadley cells terminate typically within \(\sim\)25\degr of the Equator and often well equatorward of \(\maxlat\), in which cases the cell edge sits slightly poleward of an inflection point between flat temperatures equatorward and sharply increasing temperatures poleward.  Conversely, when the forcing is supercritical from the equator to \(\maxlat\), the cells extend always into the direct vicinity of \(\maxlat\), yielding under polar maximal forcing a single, global-scale (though not quite pole-to-pole) cell even at Earth's rotation rate.  In all cases, the cells span roughly as far into either hemisphere, in reasonable agreement with the \(\urce>\uamc\) condition given values of \(\ascentlat\) and \(\uamc\) fields diagnosed from the simulation output.

\section{Discussion}
\label{sec:disc}

How relevant are these results based on dry models to moist atmospheres?  Insofar as convective quasi-equilibrium holds, \eqref{eq:e95-crit-theta} is valid in moist atmospheres.  But, as noted by E95, the irreversible fallout of precipitation in convecting towers creates an asymmetry between the ascending and descending branches of the circulation; the dry adiabatic lapse rate in the dry, descending branch due to radiative cooling-generated descent generates a stable cap over the underlying boundary layer, such that the free troposphere and boundary layer quantities decouple \citep{zheng_response_1998}.  This cannot be addressed simply via a change of variables, and \eqref{eq:e95-crit-theta} can only hold for those columns wherein such de-coupling does not occur.

Recently, \cite{singh_limits_2019} has presented a diagnostic relating the cross-equatorial cell edge to the condition of neutrality to slantwise convection \citep[\cf/][]{emanuel_air-sea_1986} that is quantitatively accurate across simulations with differing planetary rotation rate.  And \cite{colyer_zonal-mean_2019} have analyzed the scalings that emerge when the opposite assumptions as usual in the AMC model are made regarding continuity at the cell edge, \ie/ that wind is continuous while temperature is discontinuous; they find this new variant to be preferable in some ways for large \(R\) cases with planetary-scale cells.

Under purely adiabatic stratification as in the RCE state, the cells would port no heat as they overturn, implying that the cells always generate their own positive static stability via some mechanism.  \citet{caballero_axisymmetric_2008} cites two mechanisms: first, penetration of the cells into the positively stratified stratosphere, and, second, horizontal homogenization of the upper branch temperature, which is equal to the surface temperature value within the ascending branch.  Another theory that seems worth pursuing is that presented by \citet{emanuel_self-stratification_2011} for tropical cyclones: as tightly packed streamlines in the eyewall tilt from vertical to horizontal, small-scale turbulence sets in until enough static stability has been generated to relax the Richardson number to some critical value.  This same mechanism could apply to, and the corresponding formalism of \citet{emanuel_self-stratification_2011} adapted for, axisymmetric Hadley cells.  In addition to its influence on the static stability, this could potentially form a minimal theoretical model for the angular momentum mixing across upper branch streamlines noted by many authors \citep[\eg/ HH80, LH88,][]{adam_global_2009}.

In our simulations, the Hadley cells roughly conserve angular momentum in the sense that individual streamlines are nearly parallel with individual angular momentum contours in the free troposphere.  But the true theoretical AMC circulation has a \emph{single} value of \(M\), namely the planetary value at \(\ascentlat\), whereas the simulated cells feature steadily decreasing \(M\) values moving poleward.  Undoubtedly this non-uniformity relates to the finite width of the simulated ascending branch and resulting turbulent momentum mixing, as well as to momentum advection within the ascent branch.  It could\ be useful to use an alternative \(\hat\theta\) profile that gives rise to such behavior, such as the solution from the ``1 1/2'' layer model of \citet{adam_global_2010}, and use it in the place of the true \(\hat\theta\amc\) field in the equal-area model.

% The conditions of Hide's theorem hold regardless of how nearly the resulting overturning circulation homogenizes angular momentum, requiring in axisymmetric atmospheres only that free tropospheric viscosity be nonzero.  Given the deviations from the strict AMC model of Hadley cells in nearly all primitive equation numerical simulations, it thus should not come as a surprise that the extent of supercritical forcing appears to be a better predictor of Hadley cell extent than the AMC and equal-area models thus should not be especially puzzling,

%%%%%%%%%%%%%%%%%%%%%%%%%%%%%%%%%%%%%%%%%%%%%%%%%%%%%%%%%%%%%%%%%%%%%
% ACKNOWLEDGMENTS
%%%%%%%%%%%%%%%%%%%%%%%%%%%%%%%%%%%%%%%%%%%%%%%%%%%%%%%%%%%%%%%%%%%%%
%
\acknowledgments
We thank Sean Faulk and Alex Gonzalez for discussions that informed this work, Martin Singh for useful comments on an earlier draft, and Ori Adam and Kerry Emanuel for insightful reviews.  In particular, Dr. Emanuel suggested adapting the critical Richardson number idea from tropical cyclones to Hadley cells.  S.A.H. was supported by a National Science Foundation (NSF) Atmospheric and Geospace Sciences (AGS) Postdoctoral Research Fellowship, award \#1624740.  S.B. was supported by NSF award AGS-1462544.

%%%%%%%%%%%%%%%%%%%%%%%%%%%%%%%%%%%%%%%%%%%%%%%%%%%%%%%%%%%%%%%%%%%%%
% APPENDIXES
%%%%%%%%%%%%%%%%%%%%%%%%%%%%%%%%%%%%%%%%%%%%%%%%%%%%%%%%%%%%%%%%%%%%%

\appendix[A]
\appendixtitle{Derivation of Hide's theorem in zonally varying atmospheres, \cf/ \citet{emanuel_thermally_1995}}

Consider the vorticity equation
\begin{align}
  \label{eq:vort-eq}
  \pd{\zeta}{t} &= -\mb{u}\cn_\mr{h}(f+\zeta) - w\pd{\zeta}{z} \nonumber\\
  &- (f+\zeta)\nabla_\mr{h}\cdot\mb{u} + \mb{k}\cdot\left(\pd{\mb{u}}{z}\times\nabla_\mr{h} w\right)-\mathcal{D},
\end{align}
where \(\mb{u}=(u,v)\), \(\nabla_\mr{h}\) is the horizontal divergence operator, \(\mathcal{D}\) is a damping term whose functional form is irrelevant insofar as it vanishes when \(\zeta\) vanishes (as it should), and in zonally varying atmospheres a \(+\pdsl{v}{x}\) term is added to \(\zeta\).  Suppose there exists some level at which vertical velocity vanishes at all latitudes and longitudes \citep[the tropopause being a plausible candidate, insofar as it occurs at a fixed level in RCE as argued by][]{caballero_axisymmetric_2008}.  At that level, all terms of \eqref{eq:vort-eq} featuring \(w\) are zero, and the remaining terms on the right hand side also vanish if \(\eta=f+\zeta=0\).  In that case, \(\pdsl{\eta}{t}=\pdsl{\zeta}{t}=0\), and \(\eta=0\) is a stationary point, making it impossible for absolute vorticity to evolve in time from one sign to another.  Thus, given an initial resting state (or one with sufficiently weak horizontal shears that \(\eta\) everywhere takes the sign of \(f\)), a subsequent state with \(f\eta<0\) is impossible.

The \(f\eta<0\) condition has two additional physical implications.  First, it marks the onset of symmetric instability \citep{stevens_symmetric_1983}.  Second, in a purely inviscid atmosphere, it is impossible to change the sign of the Ertel potential vorticity \(\alpha\mathbf{\eta}\cdot\nabla\theta\), where \(\alpha\) is specific volume, \(\mathbf{\eta}\) is the absolute vorticity vector, and the \(\nabla\) operator is three dimensional.  Accordingly, absolute vorticity cannot change signs (unless the flow was to  generate unstable stratification).  Finally, note that the \(f\etarce<0\) condition, since it refers to the RCE state, is distinct from the argument by \cite{tomas_role_1997} that inertial instability controls the location of the ITCZ, since the latter refers to the dynamically equilibrated state.

\appendix[B]
\appendixtitle{Derivation of \(\eta=0\) location under \cite{lindzen_hadley_1988} forcing}
\eqref{eq:u-rce-lh88} implies \(M\rce<0\) where
\begin{equation}
  \label{eq:lh88-no-sol}
  \frac{2R}{1+2R}>\frac{\sinlat}{\sin\maxlat}.
\end{equation}
The left-hand side of \eqref{eq:lh88-no-sol} is at most unity, while the right-hand side is \({<}1\) for \(0<\lat<\maxlat\), guaranteeing some finite latitude range in the summer hemisphere for which no RCE solution exists.  For very large \(R\), the left-hand side approaches unity, such that the \(M\rce<0\) constraint is violated essentially all the way to \(\maxlat\), and the circulation is certain to extend at least to the vicinity of \(\maxlat\).

For the \(f\etarce<0\) condition, the absolute vorticity field corresponding to \eqref{eq:u-rce-lh88} is
\begin{align}
  \label{eq:abs-vort-lh88}
  \eta_\mr{rce,lh88} &= \Omega\sqrt{1+2R\left(1-\frac{\sin\maxlat}{\sinlat}\right)}\nonumber\\
  &\times\left(2\sinlat-\frac{\cos^2\lat}{\sin^2\lat}\frac{R\sin\maxlat}{1+2R\left(1-\frac{\sin\maxlat}{\sinlat}\right)}\right).
\end{align}
Evaluated at \(\maxlat\), this becomes
\begin{equation}
  \label{eq:abs-vort-lh88-at-maxlat}
  \eta_\mr{rce,lh88}(\lat{=}\maxlat) = 2\Omega\sin\maxlat\left(1-\frac{R}{2}\frac{\cos^2\maxlat}{\sin^2\maxlat}\right),
\end{equation}
which shows that \(f\etarce(\lat{=}\maxlat)<0\) if \(\tan^2\maxlat<R/2\), or \(\maxlat<\sqrt{R/2}\) in the small-angle limit.  For the original LH88 case with \(\maxlat=6^\circ\approx0.1\)~radians and \(R\approx0.1\), it follows that \(\sqrt{R/2}\approx0.2\), and thus the \(f\etarce<0\) condition is met at \(\maxlat\), as shown in Figure~\ref{fig:schematic-lh88}.  Figure~B1 shows \eqref{eq:abs-vort-lh88-at-maxlat} as a function of \(\maxlat\), both with and without the small angle approximation.  As the forcing maximum moves poleward, a larger \(R\) is required to ensure the circulation extends at least to \(\maxlat\) \citep{guendelman_axisymmetric_2019}.

\begin{figure}[t]
  \centering\noindent
  \includegraphics[width=19pc,angle=0, trim={0 0cm 0 0cm}, clip]{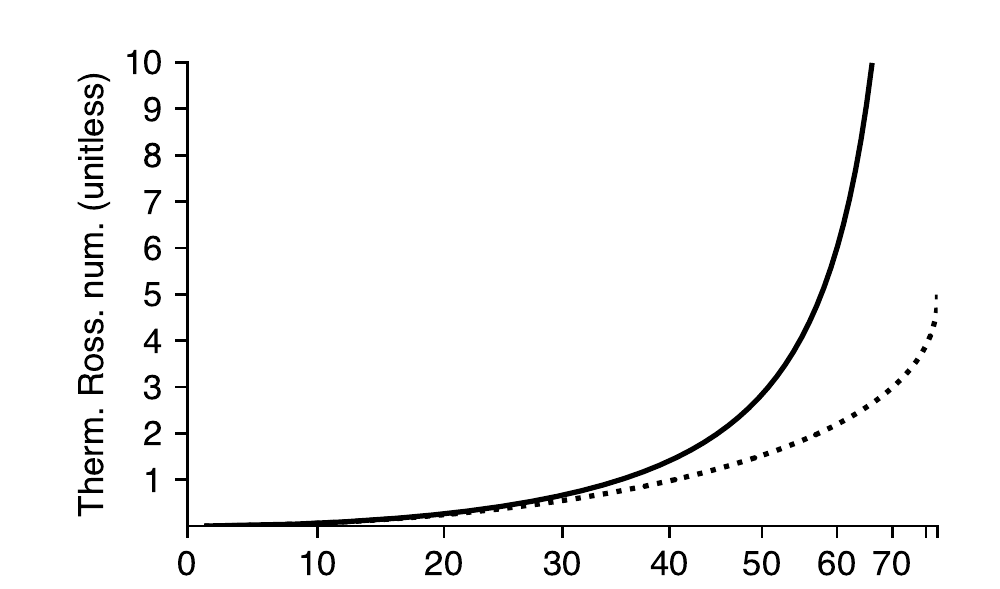}\\
  \appendcaption{B1}{Values of (horizontal axis) \(\maxlat\) and (vertical axis) thermal Rossby number for which RCE absolute vorticity at \(\maxlat\) corresponding to the LH88 forcing profile is zero.  The solid black curve is the full solution, and the dotted black curve is the small-angle limit.  Values above/to the left of the solid curve correspond to \(f\etarce<0\) at \(\maxlat\), thereby ensuring a circulation that extends to at least \(\maxlat\).}
  \label{fig:eta-lh88-phase-diagram}
\end{figure}

\appendix[C]
\appendixtitle{Dissipative processes in the idealized dry GCM}

The three dissipative processes are \(\nabla^8\) hyperdiffusion to represent subgrid-scale dissipation, quadratic damping of winds within the planetary boundary layer to represent surface drag, and vertical diffusion in the free atmosphere to suppress symmetric instabilities that otherwise cause the model to crash.  The quadratic drag formulation is \(\pdsl{\mathbf{u}}{t}=\dotsb-k(\sigma)\left|\mathbf{u}\right|\mathbf{u}\), where \(\mathbf{u}=(u,v)\), is the horizontal wind vector, \(|\mathbf{u}|=(u^2+v^2)^{1/2}\) is the horizontal wind speed, and \(k(\sigma)\) is the drag coefficient, which takes its maximal value at the surface and decreases linearly in the model's vertical sigma coordinate (\(\sigma\equiv p/\ps\), where \(\ps\) is the spatiotemporally varying surface pressure) to a value of zero at \(\sigma_\mr{b,top}\), the prescribed boundary layer top.  The planetary boundary layer top is at 0.85 to 0.7 and its drag coefficient is \(5\times10^{-6}\)~m\inv{}.  Free atmospheric viscosity is formulated as standard as vertical diffusion, such that \(\pdsl{\mathbf{u}}{t}=\dotsb+\pdsl{}{z}(\nu\pdsl{\mathbf{u}}{z})\) for zonal and meridional momentum and analogously for temperature with Prandtl number unity.  This is turned on only at model levels above a fixed height of 2500~m, a slightly different boundary layer top criterion than the fixed sigma level used by the boundary layer drag scheme.

Rather than the commonly-used uniform \(\nu\) \citep[\eg/ HH80, LH88, PH92,][]{bordoni_regime_2010}, the model uses a mixing length formulation:
\begin{equation*}
  \nu=l^2_\mr{mix}\left(1-\frac{\mr{Ri}}{\mr{Ri}_\mr{crit}}\right)^2\dfrac{|\Delta\mathbf{u}|}{\Delta z},
\end{equation*}
where \(l_\mr{mix}\) is the mixing length (a global constant), \(\mr{Ri}\) is the bulk Richardson number, \(\mr{Ri}_\mr{crit}\equiv 0.25\) is a critical Richardson number above which free atmospheric diffusion does not occur, \(\Delta\) denotes differences between adjacent model levels, and \(|\Delta\mathbf{u}|=((\Delta u)^2+(\Delta v)^2)^{1/2}\).  The bulk Richardson number is defined conventionally, \(\mr{Ri}=g\Delta\theta\Delta z/(\theta|\Delta\mathbf{u}|^2)\).  Under this formulation, the diffusivity increases with the vertical shear of the horizontal wind speed and decreases with the static stability \(\pdsl{\theta}{z}\).  We have experimented with a range of mixing length values in a subset of the simulations in order to find the lowest value in which the model integration runs successfully; that value is 15~m in all simulations except the most strongly forced simulation at Earth's rotation rate, which required a value of 30~m.

%
% Use \appendix if there is only one appendix.
%\appendix
% Use \appendix[A], \appendix}[B], if you have multiple appendixes.
%\appendix[A]
%% Appendix title is necessary! For appendix title:
%\appendixtitle{}
%%% Appendix section numbering (note, skip \section and begin with \subsection)
% \subsection{First primary heading}
% \subsubsection{First secondary heading}
%% Important!
%\appendcaption{<appendix letter and number>}{<caption>}
%must be used for figures and tables in appendixes, e.g.,
%\begin{figure}
%\noindent\includegraphics[width=19pc,angle=0]{figure01.pdf}\\
%\appendcaption{A1}{Caption here.}
%\end{figure}
% All appendix figures/tables should be placed in order AFTER the main figures/tables, i.e., tables, appendix tables, figures, appendix figures.

%%%%%%%%%%%%%%%%%%%%%%%%%%%%%%%%%%%%%%%%%%%%%%%%%%%%%%%%%%%%%%%%%%%%%
% REFERENCES
%%%%%%%%%%%%%%%%%%%%%%%%%%%%%%%%%%%%%%%%%%%%%%%%%%%%%%%%%%%%%%%%%%%%%
\bibliographystyle{ametsoc2014}
\bibliography{./references}

%%%%%%%%%%%%%%%%%%%%%%%%%%%%%%%%%%%%%%%%%%%%%%%%%%%%%%%%%%%%%%%%%%%%%
% TABLES
%%%%%%%%%%%%%%%%%%%%%%%%%%%%%%%%%%%%%%%%%%%%%%%%%%%%%%%%%%%%%%%%%%%%%
%% Enter tables at the end of the document, before figures.
%%
%
%\begin{table}[t]
%\caption{This is a sample table caption and table layout.  Enter as many tables as
%  necessary at the end of your manuscript. Table from Lorenz (1963).}\label{t1}
%\begin{center}
%\begin{tabular}{ccccrrcrc}
%\hline\hline
%$N$ & $X$ & $Y$ & $Z$\\
%\hline
% 0000 & 0000 & 0010 & 0000 \\
% 0005 & 0004 & 0012 & 0000 \\
% 0010 & 0009 & 0020 & 0000 \\
% 0015 & 0016 & 0036 & 0002 \\
% 0020 & 0030 & 0066 & 0007 \\
% 0025 & 0054 & 0115 & 0024 \\
%\hline
%\end{tabular}
%\end{center}
%\end{table}

%%%%%%%%%%%%%%%%%%%%%%%%%%%%%%%%%%%%%%%%%%%%%%%%%%%%%%%%%%%%%%%%%%%%%
% FIGURES
%%%%%%%%%%%%%%%%%%%%%%%%%%%%%%%%%%%%%%%%%%%%%%%%%%%%%%%%%%%%%%%%%%%%%
%% Enter figures at the end of the document, after tables.
%%
%
%\begin{figure}[t]
%  \noindent\includegraphics[width=19pc,angle=0]{figure01.pdf}\\
%  \caption{Enter the caption for your figure here.  Repeat as
%  necessary for each of your figures. Figure from \protect\citet{Knutti2008}.}\label{f1}
%\end{figure}

\end{document}